\documentclass[12pt]{article}
\usepackage{epsfig,amssymb}

\DeclareMathAlphabet{\mathscript}{OT1}{rsfs10}{m}{n}

\newcommand{\mat}[1]{\left(\begin{matrix}{#1}\end{matrix}\right)}
\newcommand{\eref}[1]{(\ref{#1})}
\newcommand{\sref}[1]{Subsection~\ref{#1}}

\newcommand{\cref}[1]{Chapter~\ref{#1}}
\newcommand{\bcenter}{\begin{center}}
\newcommand{\ecenter}{\end{center}}
\newcommand{\beq}{\begin{equation}}
\newcommand{\eeq}{\end{equation}}
\newcommand{\bea}{\begin{eqnarray}}
\newcommand{\eea}{\end{eqnarray}}
\newcommand{\bean}{\begin{eqnarray*}}
\newcommand{\eean}{\end{eqnarray*}}
\newcommand{\ba}{\begin{array}}
\newcommand{\ea}{\end{array}}
\newcommand{\ben}{\begin{enumerate}}
\newcommand{\een}{\end{enumerate}}
\newcommand{\bi}{\begin{itemize}}
\newcommand{\ei}{\end{itemize}}
\newcommand{\bd}{\begin{description}}
\newcommand{\ed}{\end{description}}
\newcommand{\bdiag}{\begin{diagram}}
\newcommand{\ediag}{\end{diagram}}

\def\IC{\mathbb{C}}

\def\IR{\mathbb{R}}

\def\IZ{\mathbb{Z}}

\newcommand{\be}{\begin{equation}}
\newcommand{\ee}{\end{equation}}

\def\IC{\mathbb{C}}

\def\IR{\mathbb{R}}

\def\IZ{\mathbb{Z}}

\def\re{\mbox{Re}}
\def\im{\mbox{Im}}

\newcommand{\gen}[1]{\langle #1 \rangle}

\begin{document}

\rightline{\small hep-th/0606122}

\vskip 0.5in
\centerline{\Large \bf Algorithmic Algebraic Geometry and Flux Vacua}

\renewcommand{\thefootnote}{\fnsymbol{footnote}}

\centerline{{\bf
James Gray${}^{1}$\footnote{\tt gray@iap.fr},
Yang-Hui He${}^{2,3}$\footnote{\tt hey@maths.ox.ac.uk},
Andr\'e Lukas${}^{4}$\footnote{\tt lukas@physics.ox.ac.uk},
}}
\begin{center}
${}^1${\it Institut d'Astrophysique de Paris and APC, Universit\'e de
    Paris 7, \\ 98 bis, Bd.\ Arago 75014, Paris, France} 
\vskip 0.1in
${}^2${\it Merton College, Oxford University, \\ Oxford OX1 4JD, U.K.}
\vskip 0.1in
${}^3${\it Mathematical Institute, Oxford University, \\ 24-29 St.\
  Giles', Oxford OX1 3LB, U.K.} 
\vskip 0.1in
${}^4${\it Rudolf Peierls Centre for Theoretical Physics, \\ Oxford
  University, 1 Keble Road, Oxford OX1 3NP, U.K.}
\end{center}

\setcounter{footnote}{0}
\renewcommand{\thefootnote}{\arabic{footnote}}

\begin{abstract}
We develop a new and efficient method to systematically analyse
four dimensional effective supergravities 
which descend from flux compactifications. The issue of 
finding vacua of such systems, both supersymmetric and non-supersymmetric,
is mapped
into a problem in computational algebraic geometry. 
Using recent developments in 
computer algebra, the problem can then be rapidly dealt with in a completely
algorithmic fashion. Two main results are (1)
a procedure for calculating constraints which 
the flux parameters must satisfy in these models if any given type of 
vacuum is to exist; (2) a stepwise process for finding all of the isolated
vacua of such systems and their physical properties.
We illustrate our discussion with several concrete examples, some of
which have eluded conventional methods so far.
\end{abstract}

\renewcommand{\thefootnote}{\arabic{footnote}}


\section{Introduction}

The issue of moduli stabilisation is one of the most pressing in
string phenomenology today. Recent progress in this field has resulted
in a variety of reasonably well-understood, completely stable vacua 
\cite{Strominger:1986uh,Gukov:1999ya,Giddings:2001yu,Kachru:2003aw,Grana:2005jc}.
However, these vacua, for the most part, are not physical. Two of the
greatest problems with these minima from a phenomenological standpoint
are that they do not spontaneously break supersymmetry and that they
give rise to an anti de Sitter external space.  Clearly, if we wish to
use such vacua as a starting point for building a string theoretic
description of our world this problem has to be addressed.  In the
literature, this issue is frequently resolved by employing some kind
of ``raising mechanism,'' for example, one based on the presence of
anti-branes \cite{Kachru:2003aw,Braun:2006th}, or on D-terms 
\cite{Burgess:2003ic,Achucarro:2006zf}. 
In the context of a well-controlled supergravity
descending from a string or M-theory model, there is, however, another
option. In general, such theories rich in moduli will have vacua which
exhibit spontaneously broken supersymmetry and which may be de Sitter
- even in the absence of raising of any sort.

Finding such vacua is, however, a prohibitively difficult task using 
conventional methods.
Generically, a large number of moduli fields are present in
four dimensional effective descriptions of compactified theories.
These describe 
such features of the internal space as its complex and K\"ahler structure
or the form of some vector bundle, to name but a few.
Therefore, one is confronted with potentials of supergravity theories
as complicated functions in an overwhelming number of variables.
Minimising such an expression can be beyond the reach of 
conventional techniques.

The purpose of this paper is to present a novel and efficient approach to the
systematics of finding such flux vacua. In pedagogical detail 
we provide two basic
tools which make the search for these extrema relatively easy. 
The first of these is a simple 
algorithmic process for generating constraints on the flux parameters in 
the superpotential which are necessary (and in some cases even sufficient) for 
the existence of vacua of any given type. The second tool we provide is a 
completely algorithmic way of finding all of the isolated vacua of a given 
system of interest - including non-supersymmetric vacua of the type 
described above. This tool is based upon a method for splitting up systems of
polynomial equations into multiple systems of simpler such equalities. 
Thus, we start with a set of equations which describe {\it all} of the extrema
of the potential and break these up into multiple sets of equations, where 
each of these new polynomial systems describes just {\it one} 
of the loci of extrema
of the potential (say a single isolated vacuum). In the case of isolated
vacua these new equations are so much simpler than the original expressions 
that it is found that one can solve them trivially. For
example, to entice the reader,
the following is one of the systems we discuss in later sections 
where we provide concrete examples of our methods:
\bea
K &=& -4 \log (-i(U- \bar{U})) - \log (-i (T_1-\bar{T}_1) (T_2
-\bar{T}_2) (T_3 - \bar{T}_3)), \\ \nonumber
W &=& \frac{1}{\sqrt{8}} \left[ 4 U (T_1+T_2+T_3) + 2 T_2 T_3 - T_1 T_3
  - T_1 T_2 + 200  \right] \ .
\eea
This pair of K\"ahler potential and superpotential 
has been obtained in the literature by 
compactifying M-theory on a manifold of $SU(3)$ structure \cite{Paul}.
We call the associated scalar potential, as obtained from the usual
supergravity formula, $V$. Solving for the vacua of this model directly
by solving the equations $\partial V=0$ is prohibitively difficult,
at least as far as non-supersymmetric vacua are concerned. Instead,
the method described in this paper starts by introducing a polynomial ideal
$\gen{\partial V}$, obtained from the (polynomial) numerators of the partial
derivatives of $V$. This ideal corresponds to the algebraic variety
of extrema of $V$ and can be decomposed into so-called primary ideals $P(i)$
by standard algorithms, so that $\gen{\partial V}=P_1\cap\dots \cap P_n$.
Each of these primary ideals corresponds to an irreducible variety, or,
in physical terms, a single branch of the vacuum space. Indeed, each $P(i)$
is much simpler than the original one and can be
analysed explicitly in many cases. 
In particular, the zero-dimensional primary ideals
which correspond to isolated extrema can be studied in detail using methods
of real algebraic geometry. Applying primary decomposition to 
$\gen{\partial V}$ as obtained from the above model 
(subject to the additional, simplifying constraint
$\mbox{Re}(U)=0$) leads to the following two zero-dimensional primary ideals:
\bea\label{paulsol}
\{ 3 x^2=100,t_1=2 x,t_2 = x,t_3 =x,\tau_1=0,\tau_2=0,\tau_3=0,y=0 \}
\ ,
\\ \nonumber 
\{ 9 x^2 = 500,5 t_1 = 2 x,t_2=x,t_3=x,\tau_1=0,\tau_2=0,\tau_3=0,y=0
\} \ .
\eea
Here, we have defined $T_j = \tau_j + i t_j$ for $j=1,2,3$, and $U=y+
i x$.  Thus, by breaking the equations up in this manner using the
techniques we will describe, we render the problem of finding isolated
extrema of the potential, including its stabilised vacua, trivial.
Even if cases were to exist where the simplification were not so
drastic, this still would not constitute an obstacle for us. This is because
we provide, in addition, practical algorithmic methods which can
extract all of the properties of the vacua from these equations,
without ever having to solve them explicitly.

In short, the methods we provide are practical and powerful and make
short work of finding non-supersymmetric vacua and their properties in
these flux systems.  In slightly more technical language, we propose
to re-formulate the necessary calculations arising from the
extremisation of the potential (and, indeed, extremisation problems at
large) in terms of {\bf algorithmic algebraic geometry} and {\bf
  commutative algebra}.

We will show the reader that the flux stabilisation problem
generically translates to the study of saturation and primary
decomposition of certain radical ideals in polynomial rings over
appropriate ground algebraic fields. This rephrasing is far from a
need for sophistry, but, rather, instantly allows effective
algorithms, most of which have been implemented in excellent computer
packages such as \cite{mac,sing}, to be applied. In fact, we show that
the quantities of physical interest are associated with the real roots
of complex algebraic varieties. Once the affine variety of interest
has been processed using the above complex methods, the information of
physical relevance can be extracted using real algorithmic
algebro-geometric techniques.  In particular, we make extensive use of
real root counting and sign condition routines based on the theory of
Sturm queries. These algorithms then provide us with tremendous
amounts of physical information about the vacua of ${\cal N}=1$ moduli
theories.

The methods we present find their most natural application within the context 
of perturbative stabilisation mechanisms, viz., potentials descending from 
form fluxes, torsion and non-geometric effects. In the interests of
brevity and  
clarity we therefore concentrate on such cases in this paper. 
Practically, if one is interested
in completely stabilised geometric 
vacua this would imply the consideration of models in 
type IIA or $G_2$ structure compactifications of M-theory. Non-perturbative 
effects {\it can} however be included in this type of analysis and we describe 
how this can be achieved later on in the paper.

Our approach is very much in the spirit of \cite{probe} where
a programme of systematically and algorithmically determining the
moduli space of
${\cal N}=1$ gauge theories, and in particular to look for hidden geometric
structure in the MSSM, was initiated. Here, we go one step further in our
computational capability and utilise versatile and productive
algorithms in both 
complex and real computational geometry and ideal theory. 

The paper organised as follows.
We begin in Section 2 by translating the computation of perturbative moduli 
stabilisation to one of algorithmic algebraic geometry. The problem of 
finding different vacua, SUSY, non-SUSY, Minkowski, AdS, etc., is 
classified by the type of physical questions with which one is
faced. We show in 
pedagogical detail why one is led to the study of ideals, their
radicals, as well 
as primary and saturation decompositions. Throughout we will focus on
the precise 
algorithms needed for the investigations at hand 
and how they are used in conjunction with
one another. At the end of section 2 we recover the physical classification 
presented at the start in a more mathematical context. It arises naturally in 
the process of organising the problem so that it is susceptible to the methods 
of algorithmic algebraic geometry.

In Section 3, a first example of the utility of the methods we espouse is 
provided. Using a model taken from the literature on non-geometric 
compactifications \cite{Shelton:2005cf}, 
we show how the concepts of resultants and their
multi-variate generalisation, as well as elimination-order Gr\"obner bases, 
provide us with various constraints which flux parameters must satisfy in 
such models for there to be vacua with various properties.

In Section 4, we illustrate the various methods described in Section 2 for 
algorithmically finding flux vacua and their properties. This is achieved 
by applying our methods to a sample of problems drawn from the literature, 
ranging from compactifications of M-theory to type II and heterotic 
string theories.
It is demonstrated that indeed the algorithmic methods described constitute a 
conducive path for research in the field, of diverse applicability.
Finally, we conclude in Section 5. To make the paper self-contained we
have included an 
extensive Appendix as a quick guide, first to algebraic geometry and theory 
of polynomial ideals, and second to the actual algorithms in complex and real 
geometry and commutative algebra used throughout the paper.

%
%
\section{Flux Vacua and Algebraic Geometry}
\label{method}

We wish to study four dimensional supergravity theories. 
In the context of moduli stabilisation, where the chiral superfields of 
interest are neutral under any gauge group, such theories are
specified\footnote{For 
the reader interested in charged fields we note that D-terms can
be included trivially in the discussion that follows. For the sake of
brevity we shall not, therefore, mention them further. }
by a K\"ahler potential $K$, and a superpotential $W$. 
The $K$ and $W$ which arise in such string and M-theory phenomenological 
contexts are not arbitrary. Both quantities generically take on certain
general forms which are common to all of the perturbative
stabilisation mechanisms 
currently being investigated in the literature. As such we shall concentrate
on theories with this structure.

First, we require that the K\"ahler potential be taken as
a sum of logarithms of (non-holomorphic) polynomials in the
fields. This class
of theories includes the standard form seen in the large volume and
complex structure limits of string and M-theory compactifications of
phenomenological interest.
These limits are normally considered in 
discussions of moduli stabilisation so that the use of an effective
supergravity 
is justified, and so that explicit polynomial 
formulas can be obtained respectively. We shall briefly describe how to 
extend our methods to other regions of complex structure space later.
A typical form for the K\"ahler potential of such a system is as follows:
\begin{eqnarray}\label{kahlerpotential}
K &=& -\log(S+\bar{S}) - 
 \log( d_{ijk}(T^i+\bar{T}^i)(T^j+\bar{T}^j)(T^k+\bar{T}^k) ) \\ \nonumber
&& - \log( \tilde{d}_{ijk}(Z^i+\bar{Z}^i)(Z^j+\bar{Z}^j)(Z^k+\bar{Z}^k) )
\ .
\end{eqnarray}
Here $d$ and $\tilde{d}$ are constants, which could be related to the
intersection numbers of the Calabi-Yau threefold and its mirror 
in the case of an $SU(3)$ structure compactification without intrinsic 
torsion for example. For the discussion at hand such constants will be
regarded as mere constant parameters; their origin will not be important.

Next, we must specify the superpotential $W$. 
In the same limits of large complex structure, volume and weak coupling
we again see a common form arising for the perturbative superpotentials
which are found in moduli stabilisation contexts. 
The superpotential takes the form of a holomorphic polynomial in the fields. 
This kind of superpotential includes all of the perturbative 
stabilisation mechanisms known to date: 
flux, geometrical intrinsic torsion, and non-geometric elements in the
compactification manifold.
For example, the superpotential obtained for the heterotic string 
with fluxes on an generalised half-flat manifold is given as follows
\cite{Gurrieri:2004dt,deCarlos:2005kh}:
\begin{eqnarray}\label{superpotential}
   W &=& -i(\epsilon_0 - i T^i p_{0i}) + (\epsilon_a - i T^i
  p_{ai}) Z^a + 
  \frac{i}{2} (\mu^a - i T^i q^a_i) \tilde d_{abc} Z^b Z^c\nonumber\\
             && + \frac16(\mu^0 - i T^i q^0_i) \tilde d_{abc} Z^a Z^b
  Z^c\; .
\end{eqnarray}
Here the $\epsilon$'s and $\mu$'s are parameters describing the fluxes
present in the compactified space,
while the $p$'s and $q$'s describe the intrinsic torsion. 

Non-perturbative contributions to the superpotential of course will
not take the form of a polynomial such as (\ref{superpotential}). The  
simplest implementation of the techniques we will shortly describe
requires the superpotential  
to be polynomial in the fields. Given the possibility of complete
perturbative stabilisation in some models we shall adhere
to this case for the present. Later, we shall return to the issue of  
non-perturbative contributions to the superpotential where we
shall describe how these may be accommodated within the structure we advocate.

Given the above K\"ahler and superpotentials one can proceed, 
for uncharged moduli fields, to 
construct the scalar potential from the usual formulas \cite{Wess:1992cp}. 
The scalar potential is given by:
\begin{eqnarray}\label{potential}
V = e^K \left[ {\cal K}^{A \bar{B}} D_A W D_{\bar{B}} \bar{W} - 3
  |W|^2 \right] \ .
\end{eqnarray}
As usual the $D_A$ represents the K\"ahler derivative $\partial_A +
\partial_A(K)$
and ${\cal K}^{A \bar{B}}$ is the inverse of the field space metric
\beq
{\cal K}_{A \bar{B}} = \partial_A \partial_{\bar{B}} K \ .
\eeq
Given the above-mentioned
forms of the K\"ahler potential and the superpotential, the
potential is a quotient 
of polynomials in the fields. This feature, together
with the polynomial nature of $W$, will be crucial to the
methods which we will utilise throughout this paper. We note that the 
potential can still be written as such a quotient even when raising terms 
such as those added in \cite{Kachru:2003aw} are included.

In the problem of moduli stabilisation, we are interested in finding
the extrema, and in particular the minima, of the potential
(\ref{potential}). In addition to the supersymmetric minima commonly
discussed in the literature, for which $DW=0$, this will in general
include non-supersymmetric vacua. 
These vacua can be de Sitter or Minkowski even in the absence
of D-terms or any other ``raising'' mechanisms. 
Non-supersymmetric minima of this type 
are not normally considered in the literature as even in simple 
models they are extremely difficult to find - a point to which 
we shall return shortly. The other extrema of the potential are also of some
interest. The position of 
maxima neighbouring stabilised vacua, for example, might tell us about
which set of 
cosmological initial conditions will allow the system to obtain the
stabilised configuration. 
Likewise, such information can make it possible to estimate the rate of
decay of a metastable vacuum due to tunneling.

\subsection{Classification of the Problem}
For clarity, it is expedient to classify the problem at hand into the
following subtypes, each of which shall be addressed in turn in the
ensuing sections. Let there be $n$ fields indexed by $i$, then, the
extremisation problem requires that
\begin{equation}\label{dV}
\partial_i V = 0, \;\; \mbox{for } i = 1, \ldots, n \ .
\end{equation}
We can classify the solutions to (\ref{dV}) by the amount of supersymmetry
they preserve, the value of the bare cosmological constant they dictate 
and so forth. We find it useful to define the following four subtypes:
\begin{equation}\label{classify}
\begin{array}{|c|c|}\hline
\mbox{SUSY, Minkowski} & D_i W = 0, \ \forall i, \quad W = 0 \\ \hline
\mbox{SUSY, AdS} & D_i W = 0, \ \forall i, \quad W \ne 0 \\ \hline
\mbox{NON-SUSY, Partially F-flat} & D_i W = 0, \ i = 1, \ldots, m < n
\\ \hline
\mbox{NON-SUSY, Non F-flat} & D_i W \ne 0 \ \forall i \\ \hline
\end{array}
\end{equation}

Now, recall that our potential is a rational function in the fields.
As such, the first derivatives of the potential can also be written as  
quotients of polynomials with a related denominator. Physically, we are
not interested in the solutions to the resulting equations which are
given by taking the 
denominator to infinity. These correspond to the infinite field
runaways common to these 
models. Therefore, it suffices to confine ourselves to the cases where 
the numerators of the first derivatives of the potential vanish.

In conclusion then, all the four subtypes of problems in (\ref{classify}) 
deal with the vanishing of systems of multi-variate (non-holomorphic)
polynomial equations. 
To further simplify we circumvent the issue of the presence of both
holomorphic and anti-holomorphic terms by substituting  
the expressions for the fields in terms of their real and imaginary
parts. This then reduces the problem to that of finding the real roots
of systems of 
complex polynomials. It should be noted that the problem can also be reduced 
to such a form in the presence of matter, where one would expand the potential
up to some given order in these extra fields as usual. 

\subsection{Mapping the Problem to Algebraic Geometry}
One can try and analytically solve the equations prescribed in
\eref{dV} and \eref{classify}. 
This can be quickly seen to be impossible in all 
but the most trivial cases. 
The reason for this is that, even if one of the polynomials is of 
a sufficiently low degree in a given variable to allow for an
analytical solution, when one substitutes 
this solution back into the equations to obtain a system for the
remaining variables the  degree of this system with respect to the
other degrees of freedom is increased. In 
a very small number of steps the remaining variables all appear with
degree five or 
higher and the system can not be solved. Indeed, solving systems of
multivariate polynomial equations is notoriously difficult.

Numerical techniques do not seem to fair any better. Locating the
desired minima with such methods is intrinsically difficult due to the
shallow nature of the minima and the strong features generically
present elsewhere in the 
potential. Furthermore, minima of the  types desired will in general
only appear for 
certain parameter values and this would result in a very laborious
system of trial  
and error attempts to find suitable values.
We are compelled, therefore, to seek more effective methods.

Our extrema are 
defined by the vanishing of a set of complex polynomials in the (real)
fields. Let us temporarily allow the real fields to take complex
values. This results in the submanifold of (complexified)
field space which corresponds to the extrema being defined as the
locus where a collection of holomorphic polynomials vanish. 
This is the definition of a complex algebraic variety. 
The reader unfamiliar with algebraic geometry 
is directed to the Appendices where, to make the paper as
self-contained as possible, the necessary concepts and
constructions are provided. Our
moduli stabilisation problem is then to 
{\em find the loci of real roots of a complex variety}.
As described in appendix A any given affine variety can be described
by ideals in a complex polynomial ring.
The extremisation problem of (\ref{dV}) dictates that our variety must
be defined by an ideal which is generated by the numerators of the 
first derivatives of the 
potential $V$. We shall denote this ideal by $\gen{\partial V}$.

As a technical point, multiple ideals describe the same variety. 
For example, as far as the physics is concerned, $\gen{x}$ and $\gen{x^2}$
describe the same variety, even though the ideals themselves as sets
of polynomials 
differ. To neglect such subtle scheme-theoretic differences, one can use
the so-called radical ideal, which essentially removes trivial
powers of the elements of the ideal. We denote the radical ideal
obtained from $\gen{\partial V}$ as $\sqrt{\gen{\partial V}}$. To obtain the
latter from the former, one can use a standard algorithm \cite{DGPcomp} as
implemented in \cite{mac,sing}.

Now that we have stated our problem in terms of algebro-geometrical
language we may proceed to use some of the powerful techniques 
which have been developed in that field to advance our analysis. For
clarity of notation let us first tabulate the key symbols which will
be used throughout; these will be explained in detail in Appendix A.
\subsubsection{Nomenclature}\label{nomen}
\begin{itemize}
\item $I := \gen{f_1, \ldots, f_n}$ denotes an ideal
  generated by polynomials $f_1, \ldots, f_n$. 
\item $L(I)$ denotes the variety corresponding to the
  ideal $I$ and $I(M)$ denotes the ideal corresponding to the variety
  $M$. There is reverse-inclusion in the sense that $L(I \cup J) =
  L(I) \cap L(J)$ and $L(I \cap J) = L(I) \cup L(J)$.
\item $\sqrt{I}$ denotes the {\bf radical} of ideal $I$. Hilbert's
  Nullstellensatz is the statement on the geometry-algebra
  correspondence: $I(L(J)) = \sqrt{J}$.
\item The {\bf quotient} of ideal $I$ by $J$ is denoted
  $(I:J)$. Closely related is the {\bf saturation} of $I$ by $J$,
  denoted as $(I : J^\infty)$, corresponding geometrically to the 
  sublocus of $L(I)$ which does not intersect $L(J)$.
\end{itemize}

\subsection{Techniques from Complex Algebraic Geometry}
We have now a mathematical object defining the space of extrema of the
potential: it is the
variety $L(\sqrt{\gen{\partial V}})$ corresponding to the ideal 
$\sqrt{\gen{\partial V}}$. This variety is not in general 
irreducible. Physically, this simply corresponds to the fact that the
extrema of the potential may not be connected into one piece. There may 
be isolated minima and maxima, loci of minima with flat directions and
so on.
Mathematically, this means that $\sqrt{\gen{\partial V}}$ is not a
prime ideal, but rather 
collectively contains information about all of the different
extremal loci, the union of which is the extremal variety. 
Clearly it would be useful to be able to separate out the information
about, say, lines of maxima, from that of isolated minima. 
Fortunately, a procedure exists in algorithmic algebraic geometry
which does precisely this.
\subsubsection{Primary Decomposition}
It is a theorem that any radical ideal such as $\sqrt{\gen{\partial V}}$, as we
are working over a polynomial ring over the complex numbers, is
uniquely expressible as an irredundant finite intersection of prime ideals.
Each prime ideal corresponds to an irreducible variety and 
physically represents a disconnected locus of extrema. The process of
finding these prime ideals is a heavily studied subject in algorithmic
algebraic geometry and is called {\bf primary decomposition}.
A number of algorithms have been developed to perform primary
decomposition \cite{GTZ,EHV,SY}. 
We shall make extensive use of the Gianni-Trager-Zacharias (GTZ)
algorithm \cite{GTZ} later  
on in this paper when we come to analyse examples and as such a brief
introduction  
to this is included in appendix B. This algorithm has been implemented
in \cite{sing} by GTZ and Pfister.

If we denote the prime ideal describing the $i$-th locus by $P(i)$ then, we
have the following.
\begin{equation}\label{decomposition}
\sqrt{\gen{\partial V}} = P(1) \cap P(2) \cap \ldots \cap P(k) \ .
\end{equation}
Here $k$ is the number of irreducible components of the extremal
variety - the number of different loci. The prime ideals $P(i)$ are 
in general much simpler objects than the reducible
$\sqrt{\gen{\partial V}}$. As such  
this process, even on its own, can be of considerable use in attacking 
problems of our kind. This will be seen explicitly once we move on to 
describe specific examples.

In summary, we can split up the extremisation problem of (\ref{dV}) by
performing a primary decomposition of the radical $\sqrt{\gen{\partial V}}$
of the ideal $\gen{\partial V}$. The four subtypes of the problem
according to (\ref{classify}) can, of course, be treated in the same way
and we will shortly demonstrate this concretely.

\subsubsection{Dimension and Flat-Directions}
Once we have this series of prime ideals describing the various
extremal loci for the potential of our flux system we can proceed to 
extract information about the various extrema. The extremal manifold
$L(\sqrt{\gen{\partial V}})$, using the reverse-inclusion mentioned in
\sref{nomen} and \eref{decomposition}, splits up into unions of
irreducible pieces:
\beq
L(\sqrt{\gen{\partial V}}) = L(P(1)) \cup L(P(2)) 
\cup \ldots \cup L(P(k)) \ .
\eeq 
One of the most important things to know about a given locus of 
extrema is its dimension. Our chief interest will be in 
minima which are {\em isolated in field space}; these are
fully stabilised vacua. 

For an extremum $i$ to be isolated, the dimension of the 
corresponding prime ideal $P(i)$, 
(or equivalently the dimension of $L(P(i))$) 
must be zero\footnote{The alert reader may be concerned that we are
  talking about the dimension of a complex variety when physically we
  are interested  
  in the dimension of the space of real roots. For a real root to be
  isolated it is a prerequisite that the complex dimension of the
  associated $P(i)$ is zero. If this is 
  not the case we may simply vary the real part of one of the
  unconstrained complex fields. 
  It may be the case however that a zero dimensional complex variety has
  no real roots. This is a question to which we will shortly return.}.
Physically, the piece $L(P(i))$ of the vacuum would then 
consists only of discrete points. 
In general, the $i$-th extremal locus $L(P(i))$ 
will not be zero-dimensional, and
will exhibit flat-directions, the number of these are obviously
dictated by the dimension of $P(i)$. We conclude that for all $i$,
\begin{equation}
\textnormal{Number of Flat directions of locus $i$} = \dim(P(i)) \ .
\end{equation}
Algorithms have been widely developed
for computing the dimensions of ideals. A method for testing whether
an ideal is zero dimensional, for example, is described in Appendix B.

Once we know the dimensions of the $k$ prime ideals in the 
decomposition (\ref{decomposition}) we have then obtained significant 
physical information about our system.
For example, if we were to find that none of the prime ideals are 
zero dimensional then that flux system would have no completely
stabilised vacua without flat directions,  
either supersymmetric or non-supersymmetric. If some of the prime ideals are 
indeed zero dimensional, and if we are only interested in isolated
vacua, we can then confine our 
attention to this subset of the full expansion (\ref{decomposition}).

Now, we wish to go on to answer more detailed questions. In particular, we
are interested in the following inquiries.
If an ideal is zero dimensional do any of the 
corresponding extrema correspond to real field values? 
Are the resulting isolated 
extrema maxima, minima or saddle points? Are the extrema in a well
controlled part of field space 
where we can trust the various approximations made in obtaining the
low energy effective theory  
we have been studying? Do the extrema correspond to de sitter, anti de
Sitter or Minkowski four dimensional universes? 
Are the extrema supersymmetric? To answer these questions we
need to turn to the subject of real, as opposed to complex, 
algorithmic algebraic geometry. This is the subject of the next subsection.
%
%
\subsection{Techniques from Real Algebraic Geometry}
\label{physstuff}
We now have some zero dimensional ideals $P(i)$ at hand. As discussed
above, we ultimately wish to study the real roots of our polynomial
system. We now show that
it is possible to extract the physically relevant 
information about the extrema of the potential without
ever finding the explicit location of these roots, 
in which we mostly have no interest in any event.
This situation could be compared to the use of algebraic
geometry in describing smooth Calabi-Yau compactifications. There, 
we do not know any explicit metric on the internal space yet we can
still extract much of the physically relevant information.

As a brief remark,
if we primary decompose over the complex numbers it is always 
possible to trivially solve any resulting zero dimensional
prime ideals explicitly for the 
relevant roots. The algebro-algorithmic methods described below are
still vital, however, for two reasons. First, actual implementations of 
primary decomposition algorithms normally work over the rationals where it 
is not so clear that finding explicit solutions of zero dimensional primes 
is always possible (although we have found in practice it is for these
systems -  an unexpected bonus!).
Second, these algorithms can reduce the number of costly primary decomposition 
calculations we have to perform in analysing a system. These comments will be 
illustrated concretely in later sections.

Indeed, each polynomial system $P(i)$, can be, by expanding all
of the coefficients into their real and imaginary parts (or by working
over the rationals from the start - which is what we do in practice), 
turned into a system
in $\IR[x_1,\ldots,x_n]$.
We are thus entering the realm of real algebraic geometry.
In particular, we need to know about the real roots of real 
polynomial ideals.
Much less is known about this field than about its complex cousin. 
However, it turns out that some of the few algorithms currently available
furnish us with exactly the tools we require to extract what we
wish to know.

\subsubsection{Sign Conditions and Real Root Finding} 
We will make extensive use of two kinds of algorithms \cite{realAG}. 
The first kind allows us  to compute the number of real roots of a
zero dimensional ideal (i.e., it allows us to find the number
of physical isolated extrema of our potential). 
The second allows us to compute the signs of any given set of
polynomial functions on each of the real roots of the system
\cite{realAG}, by means of a so-called Sturm query. 
A brief description of how these algorithms work is provided in Appendix C.
Both of these kinds of algorithm have been implemented in \cite{sing} by 
Tobis \cite{tobis}.

We proceed then by using the first of these algorithms to find the number of 
real isolated extrema of our potential. We then go on to use the 
second to extract the relevant physical information about these extrema.
\paragraph{Stability of the vacua: }
The double derivatives of the potential with respect to the fields for
the system specified in equations \eref{kahlerpotential} and
\eref{superpotential} take the form of quotients of polynomials which
make up the Hessian matrix $\frac{\partial^2
  V(x)}{\partial x_i \partial x_j}$. In order to check the character of the
extremum one can compute the characteristic polynomial of this Hessian
matrix (which is, in fact, a rational function) and focus on its
numerator polynomial. We can then form the ideal generated by the
characteristic polynomial and the zero-dimensional primary ideal,
describing a solution branch and perform an appropriate series of
Sturm queries on its roots. This allows you to decide algorithmically
whether the extremum is a minimum, maximum or saddle point.

Due to the effect pointed out by Breitenlohner and Freedman
\cite{Breitenlohner:1982bm} it
is necessary to determine whether these  
extrema are de Sitter, anti de Sitter or Minkowski before we can say
whether they correspond to stable vacua. If an extremum is a minimum or saddle 
point with negative cosmological constant it could still be stable. 
To discover whether this
occurs in any given case one must check the sign of a certain set of functions
\cite{Breitenlohner:1982bm}. In fact, as phrased in
\cite{deCarlos:2005kh}, the bound one needs to test, at the critical
point $x_0$ of the potential $V$, is determined by the matrix
\beq\label{BF}
\left.\left(\frac{\partial^2 V(x)}{\partial x_i \partial x_j}
- \frac{3}{2}
V(x) {\cal K}_{ij}(x) \right)\right|_{x_0} \ .
\eeq
If the eigenvalues of this matrix are all non-negative, then the AdS
minimum is stable. Indeed, for Minkowski or dS, the
positive-definiteness of the Hessian matrix $\frac{\partial^2
  V(x)}{\partial x_i \partial x_j}$ suffices for stability of the minimum.
In our case these tests again all turn out to be
quotients of polynomials and so this can be achieved with the
aforementioned algorithms.
We can therefore determine how many completely stabilised vacua the
system has.

\paragraph{Validity of the effective theory:}
For these vacua to be in a regime in which our supergravity
description is valid  
we need the values of certain fields, the size of the internal 
space for example, to be much bigger than 1 - let us say greater than
10. By checking the sign of the polynomial
$t-10$, where $t$ is the field under consideration, we can check
whether this is the case for each of our stabilised vacua.
\paragraph{Geometry of the vacua:}
The potential of the system is, as we have already pointed out, a 
quotient of polynomials. As such to deduce whether our extrema correspond
to Minkowski, anti de Sitter or de Sitter spacetimes it suffices to 
again find the sign of the numerator and denominator.
\paragraph{Supersymmetry of the vacua:}
Another important piece of information to have is whether the 
vacua are supersymmetric or not. The 
F-terms of our system, given (\ref{kahlerpotential}) and 
(\ref{superpotential}), are again rational functions and so 
we can check their sign on each of our stabilised, controlled
extrema. In particular, the algorithms described in Appendix C 
will tell us if these polynomials vanish. We can thus determine which
of the stabilised vacua are supersymmetric and which are not.

~\\

In conclusion, we can learn essentially all of the important information we
require about the vacua, both supersymmetric and  
non-supersymmetric, completely algorithmically, without ever having to
explicitly solve the system. Many of  
the interesting properties of the particle physics associated with
each vacuum can also be ascertained in this 
manner. The perturbative contributions to the masses and Yukawa
couplings in these models, for example, are rational
functions of the moduli (in appropriate limits). 
These points raised above clearly constitute a very interesting set of
questions. We will now pause our general discussions and proceed to show
how such questions may be attacked concretely.

%
%
\newpage
\subsection{Saturations and Classification Revisited} 
\label{saturationexp}
Having tantalised the readers, we now point out a caveat emptur lest
they are overwhelmed with optimism.
In practice, the above discussion has limits when pursued using the
prepackaged  implementations of the algorithms available in such
computer programs as \cite{mac,sing}. 
Indeed, naive applications of the programs often cause them to
struggle, halt, or run out of memory.

There are, luckily, various  
known tricks for avoiding this set of affairs \cite{Stillman}. These
tricks all fall under the philosophy of {\bf splitting principles} and
are concerned with splitting the problem up into more manageable 
pieces, even before passing the problem to a primary decomposition
algorithm.

One key notion in these so-called splitting principles is the idea of a 
{\bf saturation decomposition}. In this subsection, we will see how
this seemingly esoteric technique precisely adapts itself to our goal.
A more detailed definition and discussion of saturations can be found
in appendix A. Briefly, given an ideal $I$ and a polynomial $f$, 
the saturation, denoted $(I:f^{\infty})$, is equal to
\beq
sat(I,f) := (I:f^{\infty}) = \bigcup\limits_{n=1}^\infty (I : f^n) \ ,
\eeq
where each $(I : f^n)$ is the quotient of $I$ by $f^n$, which is
discussed in detail in Appendix A. The point is that the saturation
$(I:f^{\infty})$ corresponds geometrically to
the space of all
zeros of the ideal $I$ for which the polynomial $f$ does not vanish
\footnote{In fact to be precise the saturation defines geometrically the 
{\it closure} of the complement of $L(f)$ in $L(I)$. If $I$ is one dimensional
then there may be zero dimensional points in the variety associated to the 
saturation for which $f=0$ for example. In the bulk of this paper, when 
we will be interested in using saturations, our primary concern will be with
zero dimensional ideals where this subtlety does not arise. 
}.

We now follow the idea in \cite{Stillman} to utilise the
splitting principle. Suppose, for some integer $l$, the following
identity holds:
\bea
(I:f^{\infty}) = (I : f^l) \ .
\eea
In other words, at some finite $l$ the quotient has removed all powers
of $f$ from $I$. Then, we have
the following decomposition of the ideal $I$:
\bea
I = (I:f^{\infty}) \cap \gen{I,f^l} \ ,
\eea
where $\gen{I,f^l}$ is the ideal generated by $I$ together with $f^l$.
If we take the radical to neglect powers, then we have
\bea\label{split}
\sqrt{I} = \sqrt{(I:f^{\infty})} \cap \sqrt{\gen{I,f}} \ .
\eea
Geometrically, \eref{split} is the split we desire: it says that
$L(I)$ is the union of a subvariety $L(\sqrt{(I:f^{\infty})})$ where
$f$ does not vanish, with a subvariety $L( \sqrt{\gen{I,f}} )$ where
$f$ does vanish.

We pause to ask, what is a good choice of polynomial $f$, or, 
iteratively, a set of such $f$'s? 
In general, finding a non-trivial zero divisor, an element
$f$ for which $(I:f)\neq I$, can be very difficult. 
For the problem at hand, however, our supersymmetric theories
automatically provide the perfect choice!
{\em These $f$'s are simply the F-flatness conditions}. Recall that one
of our problems from \eref{classify}, the partial
F-flat case (which computationally is the most illustrative case), 
is to find the solutions to $\gen{\partial V}$ such that
$f_i = D_i W$ (or, strictly, the polynomial numerators of $D_i W$) 
vanishes only for a subset of fields $i = 1, \ldots, m <
n$. We therefore, naturally, choose 
each F-flatness equation as an $f$,
iterating from $m+1$ to $n$. Geometrically, we can write this
saturation decomposition of the vacuum manifold as:
\bea\label{satexp} 
L({\partial V}) 
&=& L(\gen{\partial V,f_1,f_2,...,f_{n}}) \cup  \\ \nonumber
&&\bigcup\limits_{i} L(\left(\gen{\partial
  V,f_1,f_2,\ldots,f_{i-1},f_{i+1},\ldots,f_{n}}:f_i^{\infty}\right)) 
\cup \\ \nonumber 
&&\bigcup\limits_{i,j} L(\left(\left(\gen{\partial V,f_1,f_2, \ldots
  ,f_{i-1},f_{i+1}, \ldots, f_{j-1},f_{j+1}, \ldots, f_{n}} :
f_i^{\infty}\right):f_j^{\infty}\right)) \cup \\ \nonumber  
&& \vdots \\ \nonumber
&& L(\left(\left(...\left(\partial V:f_1^{\infty}\right) 
  \ldots : f_{n-1}^{\infty}\right):f_{n}^{\infty}\right)) \ .
\eea
In words, what this decomposition describes is a classification of the 
different possible vacua according to how many of the F-flatness conditions
they obey. Thus the first term here is simply the supersymmetric vacuum space.
The second term is the union of all the vacuum spaces for which only one of 
the F-flatness equations is disobeyed, and so on. Once one has broken up the 
problem in this manner one can go on to apply the analysis discussed in 
previous subsections.

Therefore, this decomposition is physically intuitive, and
natural from the point of view of the theory of ideals, as well as
being practically useful. The classification \eref{classify}
corresponds precisely to \eref{satexp}.
The Minkowski vacuum, for example, would
be a subset of the first term, given by $L(\gen{\partial
  V,f_1,\ldots,f_n,W})$,
where the superpotential $W$ vanishes in addition to all of the F-flatness
conditions.
Here, a further simplification can be made; indeed,
F-flat configurations are automatically extrema of the potential in 
supersymmetric systems. Thus, the Minkowski vacuum is then
$L(\gen{f_1,\dots,f_n,W})$, 

If we wish to study a given type of vacuum - be it partially F-flat, 
non-F-flat or completely F-flat, all we have to do is to perform 
the associated saturation decomposition in \eref{satexp}. 
Working with each of these pieces is much more 
tractable than working with $\gen{\partial V}$ in its entirety. Indeed, some 
information can be extracted immediately after forming these saturations. For 
example, if a given piece in the saturation decomposition
has a dimension of $-1$ (this is the
convention that the system has no roots) then the associated set of 
vacua are absent in the model under consideration. 

We have come full circle and, in the course of setting up 
a practical method for finding minima, have recovered the physical
classification (\ref{classify}) in the more mathematical context of 
(\ref{satexp}).
We shall stop our general discussion here. In the following sections,
we will
address each of the subtypes discussed in (\ref{classify}), by
illustrating with actual examples taken from
string and M-theory phenomenology. 
In these specific
examples we will find that our method is indeed powerful.
Primary decomposition breaks the original extremely complicated sets
of polynomial equations up into more manageable  
pieces. The prime ideals containing the completely stabilised vacua are
so much simpler than the full system that they can be often solved
explicitly - thus furnishing us with a complete knowledge of the vacua
we find. 

Let us then proceed to analyse various parts of this 
expansion for a variety of models. Our aim in doing this will be to illustrate 
the power of this methodology, as well as to see what general statements can
be extracted in each case.

%

\section{The SUSY Minkowski Case and Constraints on Flux}
Let us begin with the case of supersymmetric Minkowski vacua.
Here, we are solving for the vanishing of the
superpotential and its derivatives. In this case, some general theory
can be developed and general, necessary and sufficient, conditions on
the fluxes
for the existence of such vacua can be derived. Similar constraints can be 
derived in the other cases but in those instances these are only necessary
conditions. Necessary conditions for the existence of non-supersymmetric 
Minkowski minima in supergravity have also been given in 
\cite{Gomez-Reino:2006dk}.

\subsection{Resultants and Diophantine Equations}
Before embarking on a full
discussion, let us see what happens if there were only a single
(complex) field. That is, $W$ is a degree $n$ polynomial
of a single variable $x$ with integer coefficients determined by the
values of the fluxes. We are therefore solving the system
\bea
W(x) &=& a_0 + a_1 x + a_2 x^2 + \ldots + a_n x^n = 0 \\ \nonumber
W'(x) &=& a_1  + 2a_2 x + \ldots + n a_n x^{n-1} = 0 \ .
\eea
Already, one can learn quite a lot. We know that two univariate
polynomials have common zeros iff their resultant vanishes
\cite{Harris}. Therefore, we require that
\be\label{resultant}
\ba{l}
\mbox{res}(W(x),W'(x)) = \\ \\
\det\mat{
a_n & a_{n-1} & a_{n-2} & a_{n-3} & \ldots & a_1 & a_0 & 0 & \ldots& 0
\cr
0 & a_{n} & a_{n-1} & a_{n-2} & \ldots & a_2 & a_1 & a_0 & 0 & \ldots
\cr
\vdots & & &\vdots&&&&&&\vdots \cr
\vdots & & &(n-1) \mbox{ times}&&&&&&\vdots \cr
\vdots & & &\vdots&&&&&&\vdots \cr
n a_{n} & (n-1) a_{n-1} & (n-2) a_{n-2} & (n-3) a_{n-3}
& \ldots & 2a_2 & a_1 & 0 & \ldots& 0
\cr
0 & n a_{n} & (n-1) a_{n-1} & (n-2) a_{n-2} & \ldots & 3 a_3 & 2a_2 & a_1
& 0 & \ldots
\cr
\vdots & & &\vdots&&&&&&\vdots \cr
\vdots & & &(n) \mbox{ times}&&&&&&\vdots \cr
}\\ \\
=0 \ .
\ea
\ee
In general, the resultant of an order $m$ polynomial with an
order $n$ one is homogeneous of degree $m+n$ in the coefficients. In
other words, for our case, the determinant in \eref{resultant} is a
polynomial in the $a_i$, of homogeneous degree $2n-1$. This is easy
to see. Each element in the matrix in \eref{resultant} 
is either $0$ or one of the coefficients. Each term in the determinant,
when expanded, receives one factor from each column. Any non-vanishing term
then has the same degree as the diagonal term, which is
$a_n^{n-1} a_1^{n}$, of degree $2n-1$.
This seemingly trivial observation has interesting consequences. It
dictates that the resultant vanishes, if and only if the coefficients
satisfy a homogeneous Diophantine equation.

 Now, recall that in the
general problem of studying the critical points of the (ordinary)
potential there are holomorphic and anti-holomorphic fields in our 
defining polynomials and we
needed to expand them into their real and imaginary components and
look for real roots corresponding thereto. However, our Minkowski
problem is simpler in that we need only studying the vanishing of the
(holomorphic) superpotential and its derivatives, and it suffices to
find complex roots of a purely holomorphic polynomial system as
above. Therefore, we can conclude that {\it a Minkowski vacuum exists iff
  the resultant, a homogeneous Diophantine equation in the fluxes,
  vanishes.}
Of course, there is nothing to guarantee that such vacua would be physical 
in the sense that the values of the 
real parts of the superfields would be large and 
so forth. To check whether this is the case one would have to utilise the 
methods detailed in the ensuing section.

As an illustration, let us present the resultant explicitly
for some small values of $n$:
\be\ba{|c|c|c|}\hline
n & \mbox{resultant} & \mbox{degree} \\ \hline
1 & a_1 & 1  \\ \hline
2 & a_2\,\left( -a_1^2 + 4\,a_0\,a_2 \right) & 3 \\ \hline
3 & a_3\,\left( -{a_1}^2\,{a_2}^2 + 4\,{a_1}^3\,a_3 -
     18\,a_0\,a_1\,a_2\,a_3 +
     a_0\,\left( 4\,{a_2}^3 + 27\,a_0\,{a_3}^2 \right)  \right) & 5
\\ \hline
4 &

\ba{l}
a_4\,\left( -27\,{a_1}^4\,{a_4}^2 +
     {a_1}^3\,\left( -4\,{a_3}^3 + 18\,a_2\,a_3\,a_4 \right)  -
     2\,a_0\,a_1\,a_3\,\left( -9\,a_2\,{a_3}^2 \right. \right.\\\left. \left. + 40\,{a_2}^2\,a_4 +
        96\,a_0\,{a_4}^2 \right)  +

     {a_1}^2\,\left( {a_2}^2\,{a_3}^2 - 4\,{a_2}^3\,a_4 -
        6\,a_0\,{a_3}^2\,a_4 \right. \right. \\ \left. \left. 
+ 144\,a_0\,a_2\,{a_4}^2 \right) +
    a_0\,\left( -4\,{a_2}^3\,{a_3}^2 + 16\,{a_2}^4\,a_4 +
        144\,a_0\,a_2\,{a_3}^2\,a_4 \right. \right. \\ \left. \left.
- 128\,a_0\,{a_2}^2\,{a_4}^2 +
        a_0\,\left( -27\,{a_3}^4 + 256\,a_0\,{a_4}^3 \right)  \right)
     \right)
\ea & 7 \\
\hline\ea\ee
It would be interesting to study the solutions to such Diophantine
equations. The foundational work on this subject is laid out in
\cite{GKZ}, with some recent surveys and results in
\cite{dioph,dioph2}. 

\subsection{Multi-variate Resultants and an Example}
We have discussed the univariate situation above. What about the 
general case where there is more than one variable?
 In multivariate examples the equivalents of the resultant of the
previous subsection can be computed
algorithmically using an elimination order Gr\"obner basis 
\cite{cox,DL}. 
In other words, there is a systematic method of
eliminating variables stepwise from an ideal, just like Gaussian
elimination for linear systems. A description of the algorithm for calculating
a Gr\"obner basis in the lexicographic ordering - which is an example of 
an elimination ordering - is provided in Appendix A.
This elimination, in the uni-variate
case, produces the resultant discussed in the previous subsection.

Algebraically this process takes the intersection $I \cap
\IC[X_1,...,X_n]$, of the original ideal
$I \subset \IC[X_1,...,X_n,a_1,...,a_m]$ (where the $X$'s are the
variables and the $a$'s the parameters in the original problem) with
the ring $\IC[X_1,...,X_n]$ of variables to be eliminated.
Geometrically, this 
simply corresponds to the projection of the original ideal on to the subspace
of the space described by the original ring where the eliminated
variables vanish. 
The resultant conditions on the $a$'s are then clearly necessary and sufficient
for the existence of a root of $I$ for some value of the $X$'s.

Thus, even in the multivariate case, constraints on the fluxes which
are necessary and sufficient conditions for the existence of supersymmetric 
Minkowski vacua can still be found. In some cases these can be quite compact in
form. In others, however, the resulting constraint equations can be quite 
appreciable in size, as we shall see in a concrete example now. This
constraint on the practicality of resultants of multivariate systems
above a certain level of complexity is well known \cite{GKZ}.

Let us illustrate with a concrete example from the literature.
Take equation (2.6) of \cite{Shelton:2005cf}, which presents
a non-geometric flux superpotential of the form
\begin{eqnarray} \label{sheltonW}
W & = & a_0 - 3a_1 \tau + 3a_2 \tau^2 - a_3 \tau^3\\ \nonumber
& &
 \hspace{0.2in} + S (-b_0 + 3b_1 \tau - 3b_2 \tau^2 + b_3 \tau^3)
\nonumber\\ & &
 \hspace{0.2in} + 3 U (c_0 + (\hat{c}_1 +\check{ c}_1 + \tilde{c}_1)
 \tau - (\hat{c}_2 +\check{ c}_2 + \tilde{c}_2) \tau^2 -c_3
 \tau^3), \nonumber
\end{eqnarray}
with the following constraints on the fluxes.
\bea\label{constSTW}
a_0 b_3-3 a_1 b_2+3 a_2b_1-a_3b_0 = 16 \, \,  \,\, && \\ \nonumber
a_0 c_3+a_1 (\check{c}_2 + \hat{c}_2-\tilde{c}_2)
-a_2 (\check{c}_1 + \hat{c}_1 -\tilde{c}_1)-a_3c_0 = 0 \,\,\,\, && \\ \nonumber
\ba{ccc}
\ba{rcl}
c_0 b_2-\tilde{c}_1 b_1+\hat{c}_1 b_1-\check{c}_2 b_0 & = & 0\\
\check{c}_1 b_3-\hat{c}_2 b_2+\tilde{c}_2 b_2-c_3 b_1 & = & 0\\
c_0 b_3-\tilde{c}_1 b_2+ \hat{c}_1 b_2-\check{c}_2 b_1 & = & 0\\
\check{c}_1 b_2-\hat{c}_2 b_1+\tilde{c}_2 b_1-c_3 b_0 & = & 0\\
\ea
&
\ba{rcl}
c_0\tilde{c}_2-\check{c}_1 ^ 2+\tilde{c}_1\hat{c}_1-\hat{c}_2 c_0 & = & 0\\
c_3\tilde{c}_1-\check{c}_2 ^ 2 +\tilde{c}_2\hat{c}_2-\hat{c}_1 c_3 & = & 0\\
c_3 c_0-\check{c}_2\hat{c}_1
+\tilde{c}_2\check{c}_1-\hat{c}_1\tilde{c}_2 & = & 0\\
\hat{c}_2\tilde{c}_1-\tilde{c}_1\check{c}_2
+\check{c}_1\hat{c}_2-c_0c_3 & =& 0 \ . \\
\ea
\ea \eea
There are also additional constraints which take the same form as those 
above but with the hats and checks switched around. Various useful pieces of 
algebraic processing of these constraints are provided in
\cite{Shelton:2005cf}.
These relations come from, for example, tadpole cancellation conditions 
and integrability conditions on Bianchi identities.

Finding Minkowski vacua of this system is then the problem of studying the ideal
$I = \{ W, \partial_{\tau} W, \partial_{S} W, \partial_{U} W \}$ in
the ring
$\IC(a_{0,1,2,3},b_{0,1,2,3},c_{0,1,2,3})[S,T,U]$, which is a
polynomial ring in variables $S$, $T$ and $U$ but with all fluxes
treated as parameters (formally, we call
$\IC(a_{0,1,2,3},b_{0,1,2,3},c_{0,1,2,3})$ an algebraic extension of
the ground field $\IC$). If one uses an implementation of the relevant 
algorithms in a package such as \cite{mac,sing} then it is assumed 
that none of the flux parameters vanish.
The Gr\"obner basis of $I$ in lexicographic
order then immediately gives that $I$ has negative dimension. In other
words, there are no roots in $I$. This is a quite powerful statement
without ever solving for anything, or even imposing the constraints
\eref{constSTW}: there are no Minkowski vacua for
this model, if all of the parameters are non-vanishing.

Of course, some flux parameters can vanish. So let us treat them as
variables and place $I$ in an elimination order Gr\"obner basis, and
eliminate $S,T,U$ to obtain our constraints as described above. 

The full result for the superpotential given in \eref{sheltonW} can be
obtained in  a matter of seconds\footnote{The best way to achieve this
  is to homogenise the  
problem, use a Hilbert driven global elimination order Gr\"obner
basis calculation, and then dehomogenise again at the end. See \cite{DL} for 
details.}. The result is a system of 
28 constraint equations which the fluxes must obey. 
We do not present these expressions explicitly here as they amount to 8 pages 
of expressions in this font size comprising of 6 degree 3, 12 degree
4, 8 degree 5, and 2 degree 6 polynomials.

To provide a concrete result in a presentable fashion, let us
simplify by setting, for example, all $a_{0,3}$, $b_{0,3}$ and
$c_{0,3}$ to 1. Indeed,
there are still many solutions of \eref{constSTW} with this choice. Now,
treat $I$ as an ideal in $\IC[S,T,U, a_1,a_2,b_1,b_2,
\hat{c}_1, \check{c}_1,\tilde{c}_1,
\hat{c}_2, \check{c}_2,\tilde{c}_2]$. We again proceed to eliminate $S,T,U$ 
using an implementation of elimination orderings in \cite{mac,sing}.
We find the following constraints as necessary and sufficient for the 
existence of Minkowski vacua\footnote{As before one would have to check whether
such vacua correspond to physically acceptable field values using techniques
presented in the next section.}:
\beq\ba{rcl}\label{mink-elim}
0&=&3a_2b_1-3a_1b_2+a_2c_1-b_2c_1-a_1c_2+b_1c_2,\\
0&=&27b_1b_2^2c_1+9b_2^2c_1^2-27b_1^2b_2c_2+3b_2c_1^2c_2-9b_1^2c_2^2-3b_1c_1c_2^2\\ 
&&-27b_1^3+27b_2^3-27b_1^2c_1-9b_1c_1^2-c_1^3+27b_2^2c_2+9b_2c_2^2+c_2^3,\\
0&=&27a_1b_2^2c_1+9b_2^2c_1^2-27a_1b_1b_2c_2+9a_1b_2c_1c_2-9b_1b_2c_1c_2+3b_2c_1^2c_2 \\
&&-9a_1b_1c_2^2-3b_1c_1c_2^2-27a_1b_1^2 +27a_2b_2^2-18a_1b_1c_1-9b_1^2c_1-3a_1c_1^2-6b_1c_1^2-c_1^3 \\ 
&&+18a_2b_2c_2+9b_2^2c_2+3a_2c_2^2+6b_2c_2^2+c_2^3,\\
0&=&27a_1a_2b_2c_1+9a_2b_2c_1^2-27a_1^2b_2c_2+9a_1a_2c_1c_2-9a_1b_2c_1c_2+3a_2c_1^2c_2 \\
&&-9a_1^2c_2^2-3a_1c_1c_2^2-27a_1^2b_1+27a_2^2b_2-9a_1^2c_1-18a_1b_1c_1-6a_1c_1^2-3b_1c_1^2-c_1^3\\ 
&&+9a_2^2c_2+18a_2b_2c_2+6a_2c_2^2+3b_2c_2^2+c_2^3,\\
0&=&27a_1a_2^2c_1+9a_2^2c_1^2-27a_1^2a_2c_2+3a_2c_1^2c_2-9a_1^2c_2^2-3a_1c_1c_2^2-27a_1^3+27a_2^3\\
&& -27a_1^2c_1-9a_1c_1^2-c_1^3+27a_2^2c_2+9a_2c_2^2+c_2^3 \ .
\ea\eeq
Here, $c_1 = \hat{c}_1 +\check{ c}_1 + \tilde{c}_1$ and
$c_2 = \hat{c}_2 +\check{ c}_2 + \tilde{c}_2$.
To see, therefore,
whether there are any Minkowski vacua for the choice of flux values mentioned 
above, we need only check whether
the ideal formed by joining \eref{mink-elim} and \eref{constSTW} over
ground field $\IZ$ has dimension zero or not.

In fact, in the system specified in \eref{sheltonW} and
\eref{constSTW} the Minkowski vacua always exhibit at least one flat
direction even when present. It is easy to show that the curve
given below defines a flat direction, in the $(S,U)$ plane, of the
potential obtained from~\eref{sheltonW} for any Minkowski vacuum. 
\bea
-3 a_1 + 6 a_2 \tau_0 -3 a_3 \tau_0^2 + S(3 b_1 - 6 b_2 \tau_0 + 3 b_3
\tau_0^2) + 3 U (c_1 - 2 c_2 \tau_0 - 3 c_3 \tau_0^2) =0 \eea 
Here $\tau_0$ is the expectation value of the other modulus in the vacuum.

\vspace{0.1cm}

We would like to emphasise that constraints on the fluxes such as those given 
above can be obtained in this manner for {\it any} of the cases specified in 
\eref{classify}. 
To do this, one simply takes the relevant piece in the saturation
decomposition (treating parameters as variables) and
eliminates the fields as above.
In other words, elimination orderings can provide us with 
necessary conditions on the fluxes for any type of vacuum to exist. However,
in other cases, due to our complexification of the real field space in order 
to make the relevant polynomials holomorphic, the resulting
constraints are only
necessary and not sufficient. 
This is simply because the implied roots of the 
ideal, if the constraints are satisfied, 
could correspond to complex values for 
the real and imaginary parts of our complex scalar fields. Such roots do not 
of course correspond to physical vacua. In addition, while supersymmetric 
Minkowski extrema are always minima (with the possibility of flat directions),
other forms of extrema can be unstable and therefore not correspond to vacua.

Having discussed how constraints on fluxes can be derived using elimination 
orderings we shall now resume our main discussion. In the next subsection we
revert to the question of finding vacua in flux systems according to
the methods of Section 2.

\section{Attacking the General Problem and Finding Vacua}
With the above prelude on constraints and the Minkowski case finished, 
let us systematically address the question of finding vacua, 
including the partially F-flat and Non-F-flat cases, 
using our decomposition methods as discussed in \sref{saturationexp}.
We shall consider the full expansion \eref{satexp} in several examples in
this section.

\subsection{An Illustrative Example}
Let us completely analyse a simple first example to illustrate our
method. Suppose we had a 
four dimensional ${\cal N}=1$ supergravity theory
defined by the following K\"ahler and superpotential:
\bea
K &=& -3 \log(T_1+\bar{T}_1) -3 \log(T_2+\bar{T}_2), \\ \nonumber
W &=& - T_1^2 - T_1 T_2 - T_2^2+10 T_1+10 T_2 -100 \ .
\eea

Even this simple example results in complicated equations.
Defining $T_1 = t_1+ i \tau_1$ and $T_2 = t_2 + i \tau_2 $, the
extrema of the potential $V$ is defined by the following:
\bea \label{inveqns}
0 &=& 25 (t_1^4 + t_1^2 (500
- 280 t_2 + 37 t_2^2 - 10 \tau_1^2 - 10 \tau_1 \tau_2 - 7 \tau_2^2) +
4 t_1 (-140 t_2^2 + 7 t_2^3 \\ \nonumber 
&& + 30 (100
 + \tau_1^2 + 2 \tau_1 \tau_2) + 3 t_2 (200 + \tau_1^2 + 4 \tau_1
 \tau_2 + \tau_2^2)) - 3 (20 t_2^3 + t_2^4 - 60 t_2 
 (100 \\ \nonumber 
&& +  2 \tau_1 \tau_2 + \tau_2^2) + t_2^2 (7
 \tau_1^2 + 10 \tau_1 \tau_2    
+ 10 (-50 + \tau_2^2)) + 9 (10000 + \tau_1^4 
 + 2 \tau_1^3 \tau_2 \\ \nonumber 
&& - 100 \tau_2^2 + 2 \tau_1
 \tau_2^3 + \tau_2^4  
 + \tau_1^2 (-100 + 3 \tau_2^2)))) \; , \\ \nonumber 
0&=&
  25 (18 \tau_1^3 + 27
 \tau_1^2 \tau_2 + \tau_2 (5 t_1^2 - 60 t_2 + 5 t_2^2 - 12 t_1 (5 +
 t_2) + 9 \tau_2^2) + \tau_1 (-900 + 10 t_1 
^2 \\ \nonumber && + 7 t_2^2 - 6 t_1 (10 + t_2) + 27 \tau_2^2)) \;
 ,\\ \nonumber 
0&=& 
 -25 (3 t_1^4 + t_1^3 (60 - 28 t_2) - t_2^4 - 120 
t_2 (100 + 2 \tau_1 \tau_2 + \tau_2^2) + t_1^2 (-1500 + 560 t_2 \\
\nonumber  && 
- 37 t_2^2 + 30 \tau_1^2 + 30 \tau_1 \tau_2 + 21 \
\tau_2^2) + t_2^2 (7 \tau_1^2 + 10 \tau_1 \tau_2 + 10 (-50 +
\tau_2^2)) + 4 t_1 (70 t_2^2 \\ \nonumber && 
- 45 (100 + \tau_1^2 + 2 \tau_1 \tau_2) - 3 t_2 (200 + \tau_1^2 + 4
\tau_1 \tau_2 + \tau_2^2)) \\ 
\nonumber && + 27 (10000 + \tau_1^4 + 2 \tau_1^3 \tau_2 - 100 \tau_2^2 
 + 2 \tau_1 \tau_2^3 + \tau_2^4 + \tau_1^2 (-100 + 3 \tau_2^2))) \; 
,\\ \nonumber 
0&=& 25 (-60 t_2 (\tau_1 + \tau_2)  + 5 t_2^2 (\tau_1 +
 2 \tau_2) + t_1^2 (5 \tau_1 + 7 \tau_2) - 6 t_1 (2 (5 + t_2) \tau_1 +
 t_2 \tau_2) \\ \nonumber && + 9 (\tau_1^3 + 3 \tau_1^2 \tau_2 +  
3 \tau_1 \tau_2^2 + 2 \tau_2 (-50 + \tau_2^2))) \ .
\eea
Solving this system by conventional means is clearly impossible. 
According to our discussions, let us, instead, think of \eref{inveqns}
as an ideal $\gen{\partial V} \in \IR[t_1,t_2,\tau_1,\tau_2]$. We
perform the saturation decomposition of \eref{satexp} and present
the components thereof in Table \ref{tbl:Invented}.
We have 2 complex F-flatness 
equations: $F_{T_i} = D_{T_i} W = 0$, $i=1,2$.  In the table and the 
text below we expand these into
4 real equations and take $\re[F_{T_1}] = f_1$, $\re[F_{T_2}] = f_2$,
$\im[F_{T_1}] = f_3$ and $\im[F_{T_2}] = f_4$.

With the table we can begin our analysis.
First we break up the ideals listed to extract any zero
dimensional pieces. This part of the analysis \footnote{This ancillary
  part of the process is not required in 
the algorithmisation of the problem of finding flux vacua and so was not 
mentioned in section 2. This is simply a practical point - some initial 
splitting up of the relevant ideals in this manner can make the, already 
quick, calculations involved much faster.} is performed using 
the factorising Gr\"obner basis routine \cite{DL} as implemented in 
\cite{sing}.
Once we have a zero dimensional ideal we do not decompose it any further 
at this stage.
 Anything which is not zero dimensional, however,
is primary decomposed to check whether it contains any zero
dimensional factors.
We are thus faced with a list of zero dimensional ideals; on these we
check for two conditions that they must satisfy if they are to
describe physical extrema:
\begin{enumerate}
\item The zero dimensional ideal should have real roots;
\item The real parts of our original superfields should be greater
  than 1 when evaluated at the extrema. 
\end{enumerate}
These checks are performed using the root counting and sign
query algorithms based upon  
Sturm queries as implemented in \cite{tobis,sing} and outlined in
Appendix C.

The first condition is required because our ring variables correspond 
physically to the real and imaginary parts of the physical fields. 
The second condition is physically motivated. This kind of constraint is 
enforced in systems descending from flux compactifications so 
that the vacua concerned lie both in the large K\"ahler and
large complex structure limits. 
Large values for the real parts of the equivalent of K\"ahler moduli
in these situations are required if the
effective supergravity descriptions being used in these contexts is to be 
valid.
Large complex structure generically leads to the relevant equations
being polynomial in the fields. 
Indeed, more general cases
can  be dealt with in a similar way to
non-perturbative effects (whose inclusion will be discussed later),
at least in certain limits such as the conifold one.

\begin{table}[thb]
{\begin{center}
\begin{tabular}{|c|c|l|}\hline \hline 
Ideal & Interpretation as Vacua & Physical ? \\ \hline \hline
$\gen{f_1,f_2,f_3,f_4}$ & supersymmetric  &  Yes\\ \hline
$(\gen{\partial V,f_i,f_j,f_k}:f_l^{\infty})$& partially F-flat 
& No, $t_1 =0$ or $t_2=0$\\
where $i \neq j \neq k \neq l$& & \\
and $i,j,k,l = 1,\ldots,4$& & \\ \hline
$((\gen{\partial V,f_1,f_2}:f_3^{\infty}):f_4^{\infty})$ & 
partially F-flat & No, $t_2=0$\\
$((\gen{\partial V,f_1,f_3}:f_2^{\infty}):f_4^{\infty})$& 
partially F-flat & No, No real roots\\
$((\gen{\partial V,f_3,f_2}:f_1^{\infty}):f_4^{\infty})$& 
partially F-flat & No, No real roots\\
$((\gen{\partial V,f_1,f_4}:f_3^{\infty}):f_2^{\infty})$& 
partially F-flat & No, No real roots\\
$((\gen{\partial V,f_4,f_2}:f_3^{\infty}):f_1^{\infty})$& 
partially F-flat & Yes\\
$((\gen{\partial V,f_3,f_4}:f_1^{\infty}):f_2^{\infty})$& 
partially F-flat & No, $t_1=0$\\ \hline
$(((\gen{\partial V,f_i}:f_j^{\infty}):f_k^{\infty}):f_l^{\infty})$& 
partially F-flat & No, No real roots\\
where $i \neq j \neq k \neq l$ & & \\ 
and $i,j,k,l = 1,\ldots,4$ & & \\\hline
$((((\partial
V:f_1^{\infty}):f_2^{\infty}):f_3^{\infty}):f_4^{\infty})$ & 
non-SUSY & No, No real roots\\
\hline \hline
\end{tabular}
\end{center}}
{\caption{\label{tbl:Invented} Full saturation decomposition of the
    vacuum $\gen{\partial V}$ for the potential $V$ and F-flatness
    equations $f_i$ given in the example \eref{inveqns}.
}}
\end{table}

Examining Table \ref{tbl:Invented} we see that we can confine our
attentions to just two terms in the saturation expansion; the
two physical ones, corresponding to the supersymmetric
$\gen{f_1,f_2,f_3,f_4}$ and the partially F-flat $((\gen{\partial
  V,f_4,f_2}:f_3^{\infty}):f_1^{\infty})$ extrema; both are AdS.
Indeed, as well as simplifying the analysis this allows us to 
make quite general statements. For example, all non-supersymmetric
vacua of this system are partially F-flat with $F_{T_2}$ always being
zero in the vacuum.
We proceed to study the two physical extrema in detail.

We perform the primary decomposition of $\gen{f_1,f_2,f_3,f_4}$ using 
the algorithm due to GTZ as implemented in \cite{sing}. The result 
contains 6 factors all of which are of dimension zero. 
Of these two have 1 real root, one has 2 real 
roots and 3 have no real roots.
Of the 3 factors having real roots only the single 
factor with 2 roots is such that the real parts of the superfields are valued 
larger than 1 in the vacua. Thus the physical supersymmetric vacua of the
system are given by the roots of the following ideal:
\beq
\gen{t_1-5,t_2-5,\tau_1-\tau_2,9 \tau_2^2 -175} \subset
\gen{f_1,f_2,f_3,f_4} \ .
\eeq

Now, for the partially F-flat component, the primary decomposition of 
$((\gen{\partial V,f_4,f_2}:f_3^{\infty}):f_1^{\infty})$
contains 3 factors all of which are again zero dimensional. Of these
factors two have 2 real roots and one has no real roots at all. 
Of the two factors with real roots
there is only one root in one of the factors for which the real parts
of the superfields are both greater than 1. 
This is one of the two real roots of the 
following polynomials (the one for which $t_2$ is positive):
\beq\label{sol-eg}
\gen{t_1-t_2,21 t_2^2- 20 t_2 -900,\tau_1,\tau_2} \subset
((\gen{\partial V,f_4,f_2}:f_3^{\infty}):f_1^{\infty}) \ .
\eeq

One can ask more about the properties of the above vacua, again using
Sturm queries as described in \sref{physstuff} and Appendix C. 
We find that the non-supersymmetric vacuum described above 
is not a local minimum but a saddle point by testing
the signs of the second derivatives of the potential. Furthermore, 
the vacuum in question does not obey the Breitenlohner-Freedman 
bound \eref{BF} and so this vacuum is 
not stable. Of course, in this case the resulting ideals
that need to be considered have been rendered so simple by the decomposition 
process that one can simply find the roots of the polynomials in the
prime ideals analytically.
This is in fact generically the case in these flux vacua systems
and is simply a consequence of the fact that prime ideals 
tend to take a simple form. 

\begin{figure}
\begin{center}\begin{tabular}{cc}
\includegraphics[%
  scale=0.31]{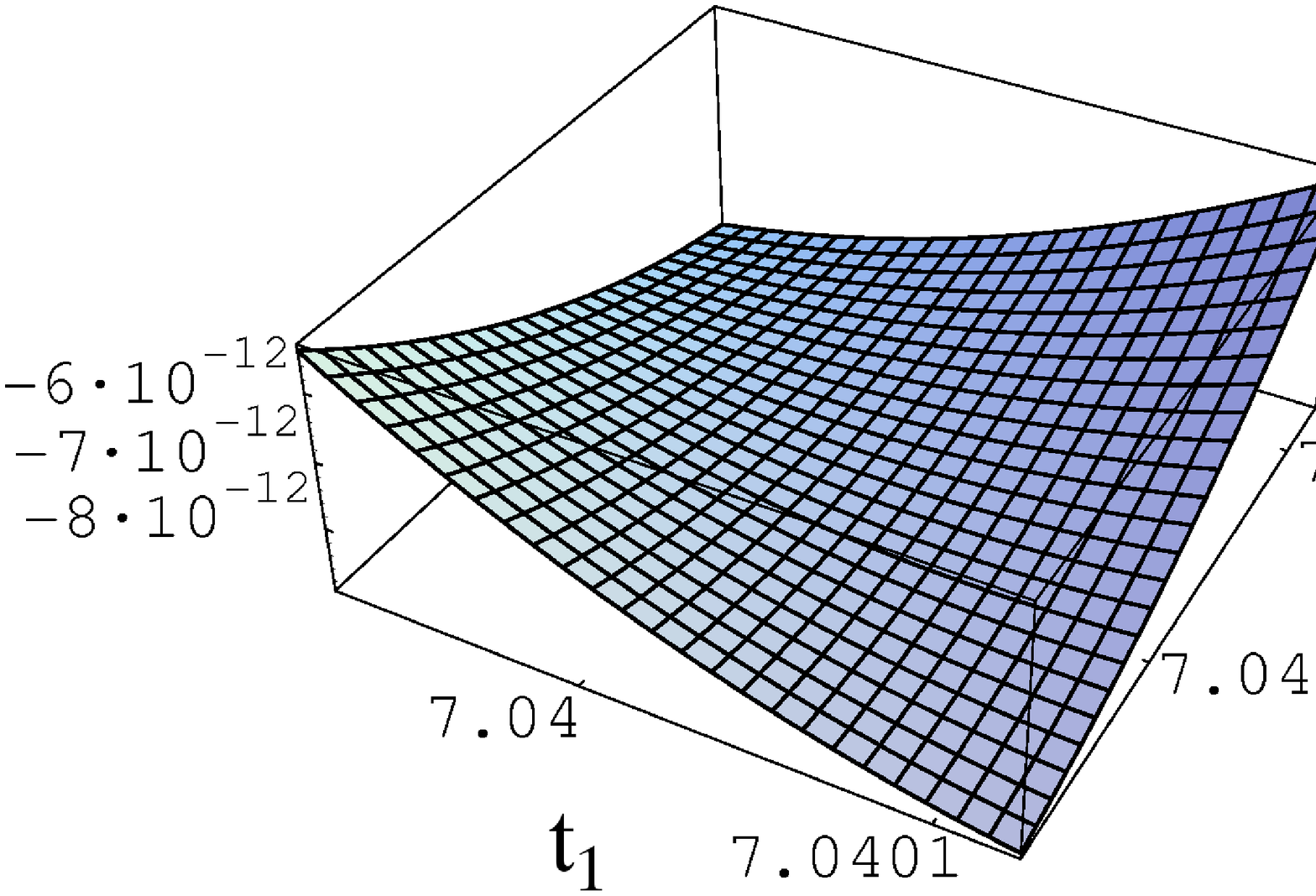} &
\includegraphics[%
  scale=0.34]{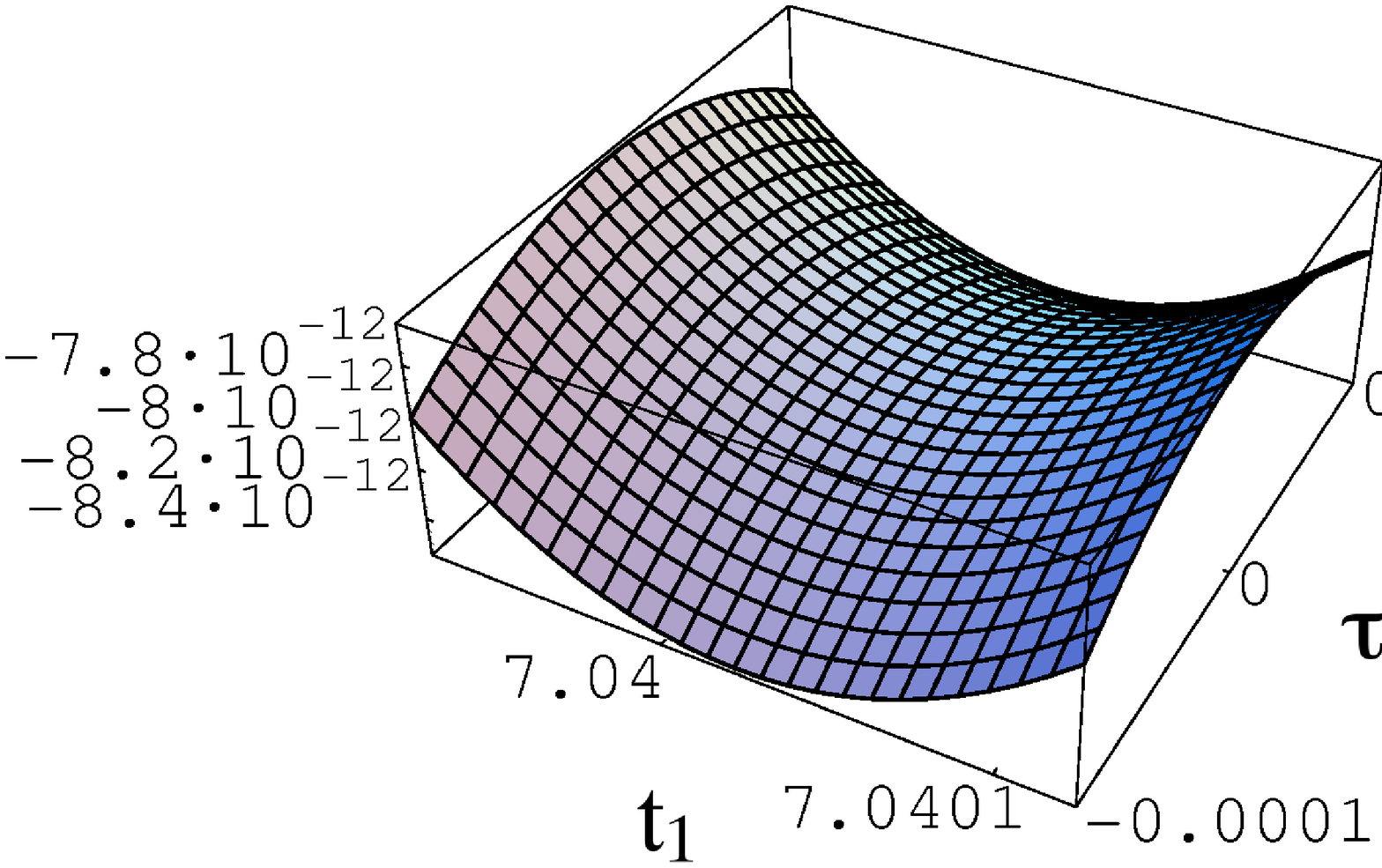} 
 \end{tabular}\end{center}
\caption{\label{f:inv} The non-supersymmetric vacuum for the
  supergravity theory specified in \eref{inveqns} for our toy
  example. The fields are $T_1 = t_1+ i \tau_1$ and $T_2 = t_2 + i
  \tau_2 $. We have plotted the potential in 
two slices through field space, viz., $t_1$-$t_2$ and $t_1$-$\tau_1$.
A shift
in the $V$ axis of $2.09279725 \times 10^{-3}$ has been performed 
so that the very shallow vacuum can be plotted effectively.} 
\end{figure}

Solving \eref{sol-eg} to find the position of the
non-supersymmetric 
vacuum we obtain $t_1=t_2=\frac{10}{21} (1
+\sqrt{190}),\tau_1=\tau_2=0$. 
Plotting the potential about this point we can therefore provide a
check that our method is functioning correctly, as is shown in Figure
\ref{f:inv}.

We have presented this example with three main goals in mind. The
first is simply to give a clear, simple example of the general
discussions given in \sref{method}. The second is to demonstrate that
this method is practical and
powerful. We reiterate that, in the system defined by \eref{inveqns},
we have found {\it all} of the isolated vacua of the system. 
It turns out in this case that there are three - two supersymmetric
and one non-supersymmetric. 
To find the non-supersymmetric vacua given above and show that it and
the F-flat solutions are the only such extrema present in the system using 
more conventional methods would be prohibitively difficult analytically. One 
would have to find all of the solutions to a system of 4 coupled quartics, even
in this simple example. Finally, the third goal is to 
show that non-supersymmetric vacua of such systems do exist, even in the 
absence of D-terms.

\subsection{Examples from String Constructions}
Having whetted the reader's appetite with our toy example,
we shall now delve into some systems which have been obtained in the 
literature in the context of string and M-theory
compactifications to four dimensions. 
Despite the complexity of the equations which appear in these contexts,
large portions of the saturation expansion (and in some
cases all of it) can still be analysed very quickly indeed.
In what follows we shall first give a simple
example from heterotic string theory.
We shall then consider an example from M-theory where all of the moduli 
of the system can be stabilised perturbatively without recourse to
non-geometric spaces.

\subsubsection{A Heterotic Example}
Let us begin with a heterotic theory compactified on one of the
$SU(3)$ structure manifolds considered in
\cite{Gurrieri:2004dt,deCarlos:2005kh}.
Of course, in a heterotic model the dilaton is unstabilised in the
absence of non-perturbative 
effects. We shall therefore just consider the stabilisation of the
analogues of the K\"ahler and complex structure moduli. 
In ignoring the dilaton in this manner the only modification to the
proceeding formulae is that the $-3 |W|^2$ term in equation
\eref{potential} becomes $-2 |W|^2$ due to a cancellation with the 
dilaton's F-term. For the K\"ahler potential and superpotential we have
\cite{Gurrieri:2004dt,deCarlos:2005kh}:
\begin{eqnarray}
K &=& - 3 \ln (T+\bar{T}) - 3 \ln (Z+ \bar{Z}) \\ \nonumber
W &=& i(\xi + i e T)+ (\epsilon + i p T) Z + \frac{i}{2} (\mu +i q
T)Z^2 + \frac{1}{6}(\rho + i r T) Z^3 \ ,
\end{eqnarray}
where $T$ is the K\"ahler modulus and $Z$, the complex structure and
$\xi, r, \epsilon, q, \mu, p ,\rho,e$ are parameters characterising the 
flux and torsion on the internal space.
These parameters satisfy the following constraint:
\begin{eqnarray}
\xi r - \epsilon q + \mu p - \rho e =0 \ .
\end{eqnarray} 
As an example let us make the following parameter 
choices\footnote{We could in principle avoid choosing 
  parameters by working over an algebraic extension of the base field
  (essentially allowing polynomials with parameter
  coefficients), as was done in the Minkowski example in Section 3. 
  Such a calculation would be expensive however. As
  such, given that the parameters in such
  models are quantised in any case, it is quicker to scan through a
  given set of values for the fluxes,  and to
  automate the following calculations. The calculations involved here are 
  sufficiently quick that this is a practical possibility.}:
\bea
\xi=-13 \; ,\; r=0 \; ,\; \epsilon=-4 \; , \; q=2 \; , \; \mu=2 \; , \;
p=1 \; , \; \rho=5 \; , \; e=-7 \ .
\eea

This gives rise to the following equations for the 
extremisation of the potential, where we have defined $T = t + i \tau$
  and $Z= z + i \zeta$:
\bea \label{heteroticeqns}
0 &=& 4t^2 z^4-12 \tau^2 z^4-25 z^6+60 \tau z^4 \zeta-48 \tau^2 z^2 \zeta^2-75 z^4 \zeta^2+120 \tau z^2 \zeta^3-4 t^2 \zeta^4-36 \tau^2 \zeta^4 \\ \nonumber &&-75 z^2 \zeta^4+60 \tau \zeta^5-25 \zeta^6+24 \tau z^4+48 \tau^2 z^2 \zeta-60 z^4 \zeta+36 \tau z^2 \zeta^2+8 t^2 \zeta^3+72 \tau^2 \zeta^3 \\ \nonumber &&- 120 z^2 \zeta^3+12 \tau \zeta^4-60 \zeta^5-108 t^2 z^2-12 \tau^2 z^2+360 t z^3-12 z^4+144 \tau z^2 \zeta-60 t^2 \zeta^2 \\ \nonumber &&- 540 \tau^2 \zeta^2-288 z^2 \zeta^2+636 \tau \zeta^3-276 \zeta^4-96 \tau z^2+56 t^2 \zeta+504 \tau^2 \zeta-192 z^2 \zeta
+ 1152 \tau \zeta^2 \\ \nonumber &&-1068 \zeta^3-196 t^2-1764 \tau^2-192 z^2+1080 \tau \zeta-1512 \zeta^2+6552 \tau-3744 \zeta-6084, \\ \nonumber
0 &=& 2 \tau z^4-5 z^4 \zeta+8 \tau z^2 \zeta^2-10 z^2 \zeta^3+6 \tau \zeta^4-5 \zeta^5-2 z^4-8 \tau z^2 \zeta-3 z^2 \zeta^2-12 \tau \zeta^3-\zeta^4 \\ \nonumber &&+ 2 \tau z^2-12 z^2 \zeta+90 \tau \zeta^2-53 \zeta^3+8 z^2-84 \tau \zeta-96 \zeta^2+294 \tau-90 \zeta-546, \\ \nonumber
0 &=& -4 t^2 z^4+4 \tau^2 z^4+25 z^6-20 \tau z^4 \zeta-16 \tau^2 z^2 \zeta^2+25 z^4 \zeta^2+40 \tau z^2 \zeta^3-12 t^2 \zeta^4 \\ \nonumber &&- 36 \tau^2 \zeta^4-25 z^2 \zeta^4+60 \tau \zeta^5-25 \zeta^6-8 \tau z^4+16 \tau^2 z^2 \zeta+20 z^4 \zeta+12 \tau z^2 \zeta^2+24 t^2 \zeta^3 \\ \nonumber &&+72 \tau^2 \zeta^3-40 z^2 \zeta^3+12 \tau \zeta^4-60 \zeta^5-108 t^2 z^2-4 \tau^2 z^2+4 z^4+48 \tau z^2 \zeta-180 t^2 \zeta^2 \\ \nonumber && -540 \tau^2 \zeta^2-96 z^2 \zeta^2+636 \tau \zeta^3-276 \zeta^4-32 \tau z^2+168 t^2 \zeta+504 \tau^2 \zeta-64 z^2 \zeta \\ \nonumber &&+1152 \tau \zeta^2-1068 \zeta^3-588 t^2-1764 \tau^2-64 z^2+1080 \tau \zeta-1512 \zeta^2+6552 \tau-3744 \zeta \\ \nonumber && -6084, \\ \nonumber
0 &=&-10 \tau z^4+16 \tau^2 z^2 \zeta+25 z^4 \zeta-60 \tau z^2
\zeta^2+8 t^2 \zeta^3+24 \tau^2 \zeta^3+50 z^2 \zeta^3-50 \tau
\zeta^4+25 \zeta^5 \\ \nonumber && -8 \tau^2 z^2+10 z^4-12 \tau z^2
\zeta-12 t^2 \zeta^2-36 \tau^2 \zeta^2+60 z^2 \zeta^2-8 \tau
\zeta^3+50 \zeta^4-24 \tau z^2+60 t^2 \zeta \\ \nonumber &&+180 \tau^2
\zeta+96 z^2 \zeta-318 \tau \zeta^2+184 \zeta^3-28 t^2-84 \tau^2+32
z^2-384 \tau \zeta+534 \zeta^2-180 \tau \\ \nonumber && +504 \zeta+624
\ .
\eea
The algebraic variety defined by these equations is reducible. First of 
all, we break up the variety according to the saturation
expansion. Despite the
fact that we are dealing with 4 coupled sextics in 4 variables 
we can calculate all but the final term (the completely non-F-flat
case) in \eref{satexp}
for the saturation decomposition extremely quickly. 
The final term takes longer to 
complete and so we will omit it in what follows. Having obtained the 
various terms in the saturation expansion we go on to study each in
turn.

We again use a mixture of the factorising Gr\"obner 
basis routine coupled with the GTZ primary decomposition algorithm, as 
implemented in \cite{sing},  
to break up the varieties. To find out which of the resulting zero dimensional 
irreducible ideals admit real roots we use the appropriate Sturm query
algorithms. We also 
study various sign conditions evaluated on these real roots and
only keep those vacua for which $\re(T)$, 
$\re(Z)>0$.
The only physical vacuum that is present is the supersymmetric vacuum which 
was found in \cite{deCarlos:2005kh}, 
there are no partially F-flat vacua in this system.
As such we shall move on to some more complicated cases with the aim
of finding some non-supersymmetric extrema. 

\subsubsection{An M-Theory Example}
Let us now look at another interesting example taken from M-theory. In
particular we would like to consider a case
where all of the moduli are perturbatively stabilised. 
We will return to the question of non-perturbative contributions to the
superpotential in the next section.

One possibility from the literature would be type IIA string theory
compactified on an
orientifold of the $\frac{T^6}{Z_2 \times Z_2}$ orbifold in the
presence of fluxes and torsion, as described in \cite{Zwirner}. 
In particular, in their subsection 5.3, these authors provide a choice of 
fluxes which results in a completely stabilised supersymmetric vacuum.

If one analyses this system using the methodology we have been
describing in this paper one instantly  
finds that $\gen{\partial V}$ for this system contains no zero
dimensional ideals at all in its primary decomposition. 
In other words there are directions in field space for which this
potential is completely flat. Once  
the presence of such a flat direction has been indicated by this
formalism it is easy to spot it explicitly  
in the potential - in this case it corresponds to a linear
combinations of some of the axions of the theory

Thus, although there is a stable supersymmetric vacuum (supersymmetric
configurations of this kind automatically obey the Breitenlohner
Freedman bound \eref{BF}), this system is perhaps
not of such strong interest for us. For example, there is no hope of
finding stable, non-supersymmetric, isolated vacua in this model. As
such we shall move on to consider another possibility.

An example of the kind we would like, better suited to our purposes, is
furnished by \cite{Paul}. These authors consider compactifying M-theory 
on the coset $\frac{SU(3) \times U(1)}{U(1) \times U(1)}$. This is a
manifold of $SU(3)$ structure. The 
resulting four dimensional supergravity theory is described by the
following K\"ahler and superpotential \cite{Paul}:
\begin{eqnarray}
K &=& -4 \log (-i(U- \bar{U})) - \log (-i (T_1-\bar{T}_1) (T_2
-\bar{T}_2) (T_3 - \bar{T}_3)) \ , \\ \nonumber
W &=& \frac{1}{\sqrt{8}} \left[ 4 U (T_1+T_2+T_3) + 2 T_2 T_3 - T_1 T_3
  - T_1 T_2 + 200  \right] \ .
\label{paulKW}
\end{eqnarray}
To give an idea of the complexity involved in a case such as this
we note that the potential takes the form:
\beq \label{paulpotential}
\ba{rcl}
 V&=& \frac{1}{256 t_1 t_2 t_3 x^4} (40000 + t_3^2 \tau_1^2 -
        400 \tau_1 \tau_2 - 4 t_3^2 \tau_1 \tau_2 + 4 t_3^2 \tau_2^2 +
        \tau_1^2 \tau_2^2 - 400 \tau_1 \tau_3 + 800 \tau_2 \tau_3 +
        \\ \nonumber && 2 \tau_1^2 \tau_2 \tau_3 - 4 \tau_1 \tau_2^2
	\tau_3  + \tau_1^2 \tau_3^2   
- 4 \tau_1 \tau_2 \tau_3^2 + 4 \tau_2^2 \tau_3^2 - 24 t_2 t_3 x^2 +
        4 t_3^2 x^2 - 24 t_1 (t_2 + t_3) x^2 \\ \nonumber && + 4 \tau_1^2 x^2 +
        8 \tau_1 \tau_2 x^2 + 4 \tau_2^2 x^2 + 8  \tau_1  \tau_3  x^2   
+ 8  \tau_2  \tau_3  x^2 + 4  \tau_3^2  x^2 + 1600  \tau_1  y -
        8  t_3^2  \tau_1  y \\ \nonumber && + 1600  \tau_2  y + 16
	t_3^2  \tau_2  y - 
        8  \tau_1^2  \tau_2  y - 8  \tau_1  \tau_2^2  y + 1600 \tau_3  y -
        8  \tau_1^2  \tau_3  y + 16  \tau_2^2  \tau_3  y - 8 \tau_1
	\tau_3^2  y  
 \\ \nonumber && + 16  \tau_2  \tau_3^2  y + 16  t_3^2  y^2 + 16
 \tau_1^2  y^2 + 
        32  \tau_1  \tau_2  y^2   + 16  \tau_2^2  y^2 + 32  \tau_1
	\tau_3  y^2 + 
        32  \tau_2  \tau_3  y^2 + 16  \tau_3^2  y^2 \\ \nonumber && 
+ t_1^2   (t_2^2 + t_3^2 + \tau_2^2 + 2  \tau_2  \tau_3 + \tau_3^2 + 
              4  x^2  - 8 \tau_2  y - 8  \tau_3  y + 16  y^2 ) +
        t_2^2   (4  t_3^2 + \tau_1^2 - 4  \tau_1  (\tau_3 + 2  y ) \\
	\nonumber && + 
              4   (\tau_3^2 + x^2 + 4 \tau_3  y + 4  y^2)) \ ,
\ea\eeq
where we have defined the component fields by
$T_j = -i t_j + \tau_j$ for $j=1,2,3$, and $U = -i x+  y$. 

To obtain the 
equations for the extrema of this potential we must then take the
derivatives of this expression with respect to all 
8 fields and set them equal to zero. The result is somewhat lengthy
and so we shall spare the reader the explicit
full set of conditions for the extremisation of this potential. To
solve these equations using normal techniques we 
would have to solve 8 coupled equations in 8 variables with each
equation involving a quotient of a fourth order and  
seventh order polynomial, clearly an impossible task (even for
packages such as Mathematica or Maple).

However, using our saturation and primary decomposition techniques,
the problem is much more tractable. Now, 
in the interests of showing the diverse manners to which our methods
can be applied, we will present a slightly 
different analysis for this system. It may be the case that one wishes
to examine vacua with certain physical  
properties besides a specific degree of F-flatness. For example,
one can ask if there are any vacua 
for any particular field values; for instance, say $y=0$.
In terms of the variety being considered this is associated with the
ideal which is generated by  
$\partial V$ and the monomial $y$. This system, which would still be
prohibitively difficult to  
solve with more conventional techniques, is well within the
capabilities of our algorithmic techniques on a desktop computer. 
The search for such vacua might
be physically motivated in many ways.
For example, one may wish certain axions in certain models to vanish in
the vacuum in order to agree with a  
small theta angle in a desired target theory. Since we are using the
power of this formalism to look at  
stable, non-supersymmetric vacua, demanding such physical inputs hold
true is now a reasonable thing to do. 

Again, using a combination of factorising standard basis, GTZ primary
decomposition and Sturm query algorithms to decompose and analyse the  
ideal $\gen{\partial V, y}$ one obtains a decomposition involving 16
factors each of which may be made up of  
numerous prime factors themselves. Many of the resulting prime factors
are zero dimensional but only two have  
real roots for which the real parts of all of the superfields take
values greater than 1.

As before, the prime ideals which we have extracted from the overall
problem to describe these isolated loci are so simple that we can
solve them explicitly to find the extrema. These turning points are
described by the following ideals (the generators of which should be
compared in complexity with the first derivatives of equation
\eref{paulpotential}): 
\bea\label{paulsolI} I_1 &:=& \gen{3
  x^2-100,t_1-2 x,t_2 - x,t_3 -x,\tau_1,\tau_2,\tau_3,y}, \\ \nonumber
I_2 &:=& \gen{9 x^2 -500,5 t_1 - 2
  x,t_2-x,t_3-x,\tau_1,\tau_2,\tau_3,y} \ .  \eea 
The simplicity of
these equations shows us how useful this procedure is. In separating
out the ideals that describe the isolated extrema in which we are
interested from all of the rest of the turning points we have vastly
simplified the discussion of the minima - in this case rendering it
rather trivial.  The physical root of $I_1$ is simply the
supersymmetric vacuum of the system. This reproduces the result found
in \cite{Paul}. The physical root of $I_2$ is an isolated extremum of
the system which is non-supersymmetric and anti de Sitter. These two
constitute {\it all} of the isolated extrema of this system which obey
the physical constraint we have imposed.
The SUSY extremum is Breitenlohner-Freedman stable while the non-SUSY
one is not.

\begin{figure}
\begin{center}\begin{tabular}{cc}
\includegraphics[%
  scale=0.34]{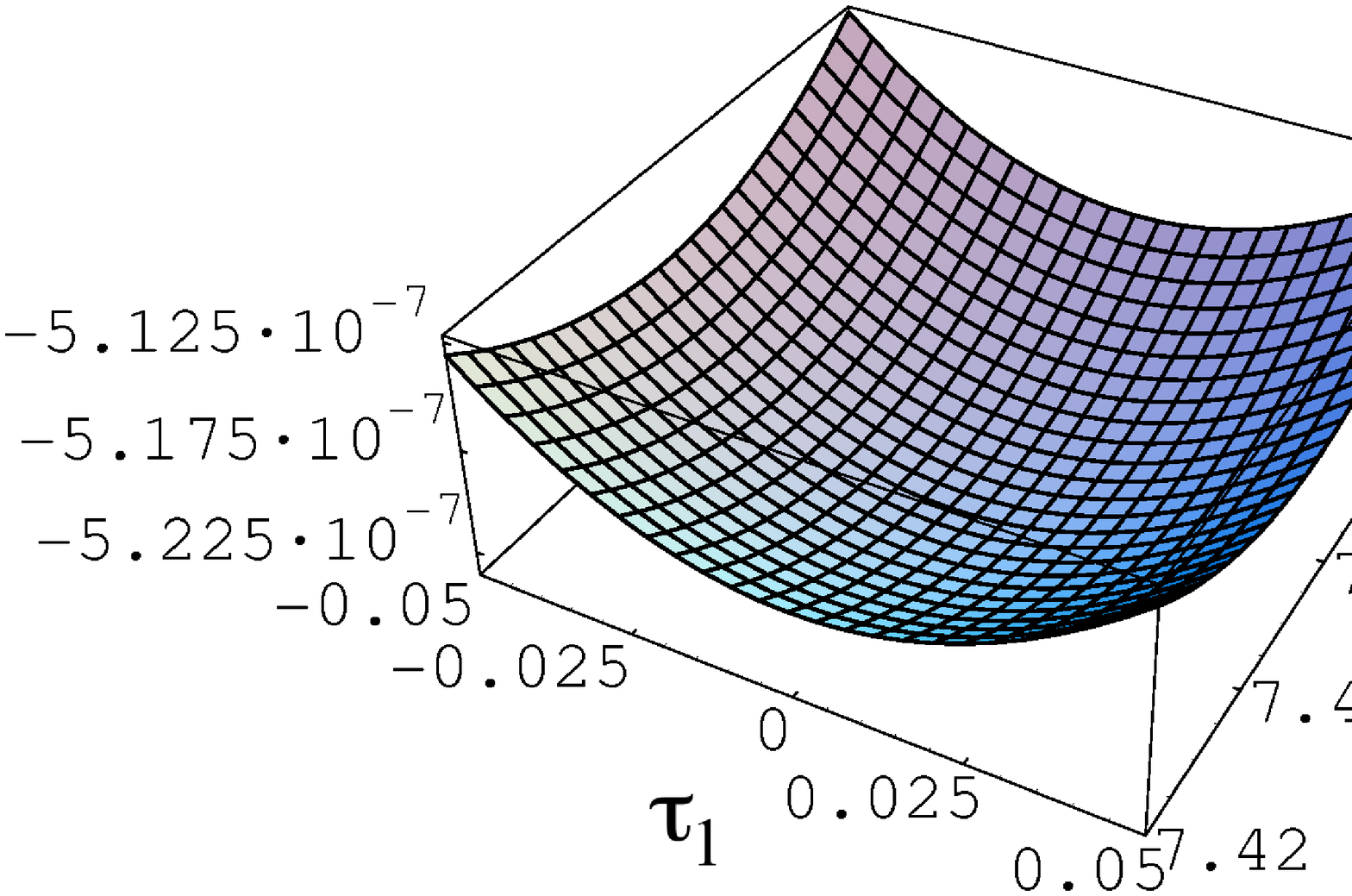} &
\includegraphics[%
  scale=0.31]{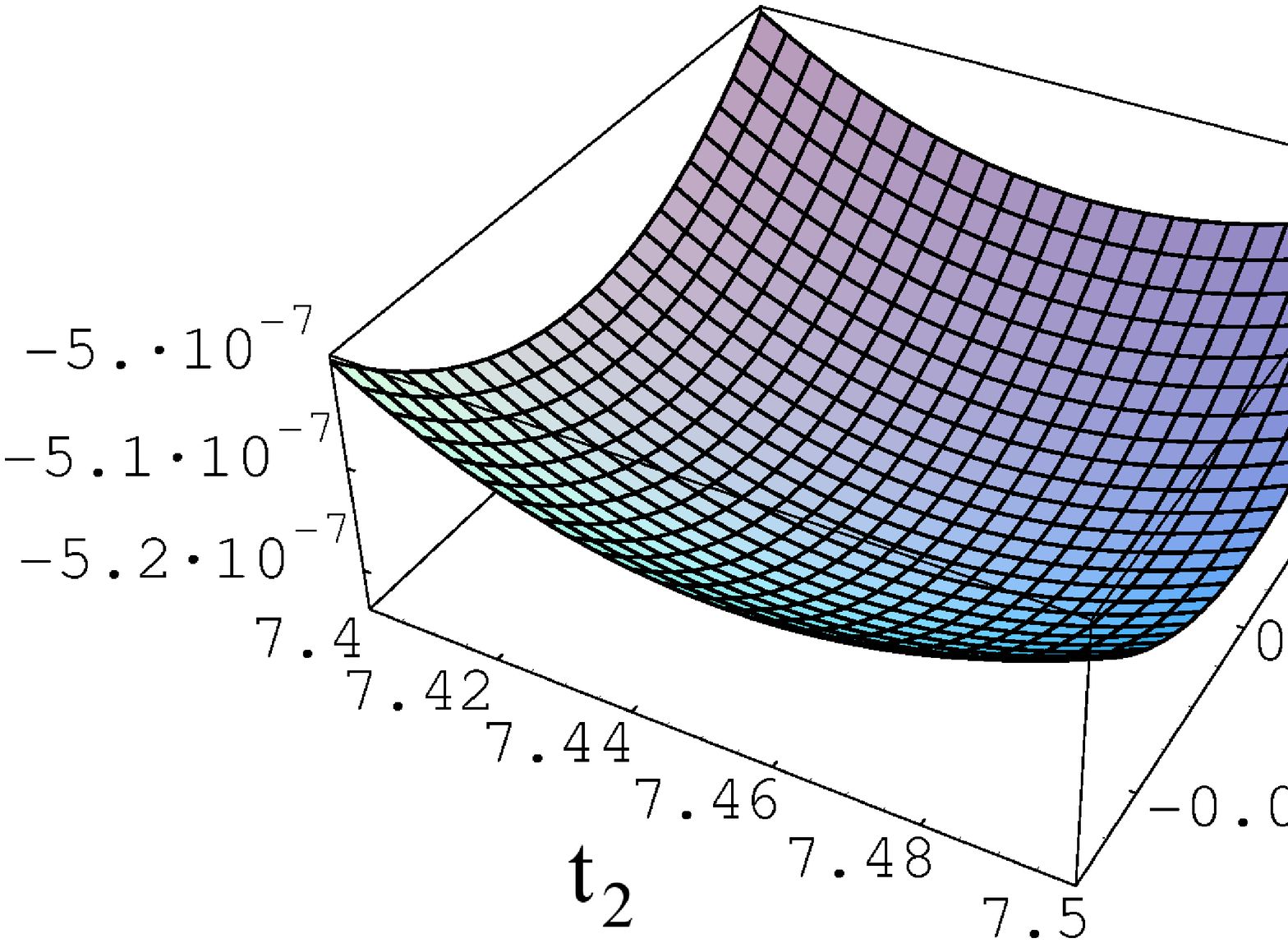}
 \end{tabular}\end{center}
\vspace{-0.3cm}
\caption{\label{theplot2} The non-supersymmetric extremum
  corresponding to the ideal $I_2$ in \eref{paulsolI}, for the supergravity
  potential specified in \eref{paulKW}.
  The fields are $T_i = -i t_i + \tau_i$ and $U = -i x+  y$; we have
  here plotted the slices in $(\tau_1, t_3)$ and $(t_2,y)$ coordinates. A shift
in the $V$ axis of $4.07 \times 10^{-4}$ has been performed so that the very 
shallow vacuum can be plotted effectively.
} 
\end{figure}
We see that the plots of Figure \ref{theplot2}
 confirm all of the features of the non-supersymmetric extremum that
our algorithmic algebro-geometric  
procedure rapidly predicted. We have also calculated a large part of the
saturation expansion \eref{satexp} for this case. 
We do not however find any interesting extrema beyond those described
above and so shall not explicitly present this analysis here.

\section{Conclusions and further work}

This paper was concerned with the problem of finding vacua of four 
dimensional supergravities describing flux compactifications.
After presenting a natural classification of such vacua we have 
provided two primary results within this context.

First, we have described a practical, algorithmic 
method for generating constraints on the flux parameters in the 
superpotentials of such systems. 
We emphasise again that these constraints can be derived 
as necessary conditions for the existence of
{\it any} given kind of vacuum. In the case of supersymmetric Minkowski vacua 
this result is even more powerful. For these special vacua the constraints
we have provided are both necessary and sufficient for the existence of
such extrema.

Second, and perhaps more importantly, we have outlined a completely 
algorithmic method to find {\it all} of the isolated vacua of such systems,
be they supersymmetric or not. In addition to the vacua themselves 
the methods we have described enable us to algorithmically find 
most of the quantities of physical interest associated with them. This includes
the degree of supersymmetry they preserve, their stability as well as 
particle physics properties such as the Yukawa couplings in the matter
sector.

What we have done is to map the extremisation problem to the language
of algorithmic algebraic geometry and, in particular, of ideal theory
and commutative algebra.
This is not simply a hypothetical discussion. Using recent advances in
computer algebra, the methods we present are 
powerful and allow us to solve, within seconds on an ordinary desktop
computer, problems which are simply impossible with conventional
techniques.
We have demonstrated in concrete examples the efficiency with which
our algorithms can find isolated non-supersymmetric
extrema in actual systems directly derived from string and
M-theoretic compactification.

One obvious extension of the work presented here would be the inclusion of 
non-perturbative elements in the superpotentials considered. There are several
ways in which one might do this. The simplest way to proceed would be
to simply  
introduce extra `dummy' variables to represent any exponential functions that 
appear. One would then have as the desired vacuum space an algebraic variety,
as described in the bulk of this paper, intersected with an
exponential equation 
- that defining the dummy variable. This would enable one to bring the full 
power of algebraic geometry to bear on the difficult part of the problem.
Another possibility would be to fix field values at some desired
values and then
solve the system to see what flux values are required to give stable vacua.
In other words, we can solve for a set of 
the parameters rather than the fields. If this is performed
carefully this will result in an algebraic variety, with the fluxes as the
variables, as the object to be analysed. This approach seems to be
more difficult
to pursue, however, due to the quantised nature of the flux parameters in these
systems.

In any event, in this paper we have restricted ourselves to perturbative 
superpotentials where the methods we have outlined find their simplest 
application. Such superpotentials can result in stabilisation of all of 
the moduli in geometric IIA and M-theory compactifications. Non-geometric 
compactifications (which give rise to a perturbative superpotential) can 
give rise to stabilised vacua in the other string theories as
well. Perturbative vacua
are interesting as they are on a somewhat firmer footing than their
counterparts
which rely on a mixing of perturbative and non-perturbative effects. One 
reason for this is that in such mixed scenarios one relies on a play
off between the two types of superpotential contribution to obtain a vacuum.
Although such playoffs are theoretically possible with the rest of the
infinite series of non-perturbative corrections being negligible, such
a situation
is dependent on, for example, a very large coefficient appearing 
in front of the exponential terms. There 
is no reason to believe that such a coefficient would arise in any given model.

As a side comment we note that all of the non-supersymmetric vacua we have 
found thus far in any model have been partially F-flat. 

Further extensions to this work are clear and numerous. As well as
the inclusion
of non-perturbative effects mentioned above 
one could consider improving the algorithms used
and their application to the problem at hand. One possible such direction of 
improvement would be to construct a method for performing the calculations over
a finite field and then separating the spurious results from the physical ones.
Gr\"obner basis calculations 
over finite fields can be {\it much} faster than those over the rationals.

Finally, pushing these methods to their natural conclusion, one could imagine
a completely automated algorithmic approach to extracting the phenomenological 
physics from four dimensional descriptions of string compactifications. 
Once a four dimensional effective theory
is derived we have shown that we can scan the vacua of the system and their 
properties algorithmically - searching for appropriate minima with which to 
describe our world. Due to the complexity of these problems
\cite{Denef:2006ad} such a 
program of research would have to be guided by physical insight. 

\section*{Acknowledgments}
The authors would like to thank Eran Palti. J.G. is grateful to the 
University of Oxford for hospitality while some of this work was being 
completed. J.G.~is supported by CNRS.
Y.H.H.~is supported by the FitzJames Fellowship at Merton College,
Oxford.

\appendix
\section*{APPENDIX}
\section{Rudiments of Computational Algebraic Geometry}
Our computations throughout this paper have relied heavily upon
techniques and algorithms in algebraic geometry, which may not be
entirely familiar to all researchers in the field. With this
appendix, which will provide a glossary on the key concepts used, we
wish that the subsequent self-contained nature of this paper may serve
the incipience of such methods into the study of flux vacua. Detailed
exposition can be found in the texts of \cite{Harris,hart}, whose
emphasis is on the theory, of \cite{cox}, on the computation, of
\cite{mac,sing,DL}, on the practically implemented algorithms, as well as
of \cite{probe}, on a
parallel application to ${\cal N}=1$ gauge theories.

\paragraph{Algebraic Varieties and Ideals: }
The problem of finding the vacua of our concern, as we stated
earlier, is the problem of finding the set $M$ of simultaneous zeros
of a system of polynomial equations in variables $x_{1},\ldots,x_{n}$. 
Such a set $M$ is an {\bf affine
  algebraic variety}. In the language of commutative algebra, in which
actual algorithms are always phrased, this set is seen as the loci of 
roots of an {\bf ideal} 
$I(M)$ in the ring $R=\IC[x_1,x_2,\ldots,x_n]$ of polynomials in $x_i$
with coefficients in $\IC$. 

Briefly, recall that a ring is roughly a
set with addition (and its inverse, subtraction) and multiplication,
but no division. Indeed, the sum, difference and product of two
polynomials remain a polynomial while the ratio does not. An ideal is
a subset, which, when multiplied by any element, remain in the
subset.
To intimate the relation between the algebraic object, viz., the ideal
$I$ and the geometric object, viz., the variety $L$, the standard
notation is to use $I(L)$ and $L(I)$ when they correspond.

To be explicit, we use $\gen{f_1,\ldots,f_k}$ to denote the ideal
of generated by the polynomials $f_i$, i.e.,
\begin{equation}
I = \gen{f_1,\ldots,f_k} = \left\{ \sum\limits_{i=1}^k  
h_i(x_1,\ldots,x_n) f_i
\right\} \subset \IC[x_1,\ldots,x_n]
\end{equation}
for polynomials $h_i$. In this notation, addition and multiplication
between two ideals is easily defined as the addition and
multiplication of all combinations of the generators. Quotients will
be defined shortly.

\paragraph{Radical Ideals: }
Next, $I(M)$ can contain more information than is physically
needed. Multiplicities in the roots describe the same set of
points. Recall the example in the text: $x=0$ and $x^2=0$ describe the
same set of points even though $\gen{x}$ and $\gen{x^2}$ are two
different ideals. This ambiguity is resolved by 
defining the radical $\sqrt{I}$ of the ideal of $I$
in a ring $R$:
\beq
\sqrt{I} := \left\{r \in R | r^n \in I \mbox{ for some } n \in \IZ_+
\right\} \ .
\eeq
The Hilbert
Nullstellensatz then states that, for any ideal $J$, the ideal $I(L(J))$
corresponding to the variety $L(J)$ whose points are determined by $J$
is equal to the radical ideal $\sqrt{J}$. In other words, the radical
ideal is the ``minimal'' ideal corresponding to the variety $M$ which
drops all the redundant information on the multiplicities of the
zeros. Thus, we can refine to the study of the radical ideal
$\sqrt{I(M)}$ corresponding to our zero-set $M$. Popular algorithms which
perform this step can be found in \cite{DGPcomp} and are implemented in
\cite{mac,sing}.

\paragraph{Primary Decomposition: }
The radical 
ideal $\sqrt{I(M)}$ may still be reducible in the sense that the variety
$xy=0$, for example, clearly consists of two irreducible components
$x=0$ and $y=0$. To obtain the elemental constituents of $\sqrt{I(M)}$ we
must then decompose it into {\bf prime ideals}, ideals $p$ for which (just
like a prime number), $ab \in p$ implies that $a\in p$ or $b\in
p$. Such a process is called primary decomposition\footnote{Strictly,
  irreducible varieties correspond to {\bf primary ideals} which are
  ideals $I$ for which $ab \in p$ implies that $a\in p$ or $b^n\in
  p$ for some integer $n$, a weaker condition than primality. However,
  since radicals of primary ideals are prime and we are already starting
  with a radical ideal, it suffices to study the stronger condition of
  prime decomposition.}.
The theory was originally due to Lasker-Noether, with the first
algorithm by Hermann. Today, it constitutes one of the most exciting
areas of research in computer algebra, with popular algorithms by
Shimoyana-Yokoyama, Eisenbud-Huneke-Vasconcelos, and
Gianni-Trager-Zacharias as implemented in \cite{mac,sing}. We shall describe
the last of these algorithms, which we have used throughout this paper, 
in some detail in Appendix B. 
We therefore have the decomposition of $\sqrt{I(M)}$ as the finite 
intersection of prime ideals $P(i)$, i.e., 
$\sqrt{I(M)} = \cap_i P(i)$.

\paragraph{Real Roots: } After decomposing into irreducible
components, one can then compute the dimension (corresponding to the
number of flat directions) of each piece $P(i)$.
A method for checking whether an ideal is zero dimensional, for example, 
is briefly described in appendix C.
In the case that the ideal $P(i)$ is
zero-dimensional, the component corresponds to no more than a
(discrete) set $S_i$ of  points.
Physically, this means that this component of the vacuum has been
completely isolated. One could determine the cardinality of $S_i$
(the number of roots of the polynomial system);
this is called the {\bf virtual dimension} of the zero-dimensional
ideal $P(i)$.
In particular, we are interested in the set of real roots, which is a
special subset of $S_i$. Algorithms have been developed to deal with real 
roots \cite{realAG}. We shall discuss this further in appendix C.

\paragraph{Quotients and Saturations of Ideals: }
During the course of our analysis we need the concepts of 
saturations and quotients of ideals. An ideal quotient of an ideal 
$I \subset R$ with respect to $f \in R$ is simply defined as follows:
\be
(I:f) := \{ g \in R | gf \in I \} \ .
\ee
In general, the quotient $(I:J)$ 
of an ideal $I$ by an ideal $J$ is the set of elements $g \in R$ such that
$g \cdot J$ is contained in $I$.
The definition of a saturation of an ideal is then a simple extension
of this idea:
\be
(I:f^{\infty}):= \{ g \in R | gf^N \in I, \textnormal{for some } N \in
\IZ_{>0}\} = \bigcup\limits_{n=1}^\infty I : f^n \ .
\ee
The second equality is
important and is the origin of the infinity in the notation:
saturation quotients out all powers of $f$.
Geometrically, this means that $L(I:f^{\infty})$ corresponds to the
subvariety of $L(I)$ for which $f \neq 0$.

\paragraph{Quotient Rings:}
The last concept that we shall require, for use in later appendices, is that 
of a quotient ring. For an ideal $I$ in a ring $R$ the quotient ring $R/I$ is 
simply defined to be the set of all elements in $R$ where two elements are
regarded as equivalent if their difference is an element in $I$.
Physically the quotient ring corresponds to the set of all
polynomial functions 
where two functions are only regarded as different when they 
take different values on the locus $L(I)$ which is defined by the
ideal $I$.

\paragraph{Gr\"obner Basis: }
The first step in almost all
algorithms in computational 
algebraic geometry is to place the generators of the ideal
of multi-variate polynomials into a
so-called Gr\"obner Basis.
This is a generalisation of Gaussian elimination for a multivariate
linear system to general polynomials.

In computational algebraic geometry, the Gr\"obner basis is determined
by (modifications and improvements of) Buchberger's
algorithm (see for example \cite{DL}). 
The Buchberger algorithm proceeds as follows. Start with an ideal $I$.
\begin{enumerate}
\item Set ${\cal G}=\mbox{generators}(I)$.
\item For any pair of polynomials $A$, $B \in {\cal G}$ form the $S$
  polynomial
(described below).
\item Reduce the $S$ polynomial with respect to ${\cal G}$.
\item If the reduction is non-zero add the result to ${\cal G}$.
\item Repeat from step 2 until all pairs of polynomials in ${\cal G}$ give $S$
polynomials which reduce to zero. ${\cal G}$ is then the Gr\"obner basis.
\end{enumerate}
In the above one needs to understand the process of reduction and what an $S$ 
polynomial is. Both of these concepts rely on the introduction of {\bf
  monomial  
orderings}. An ordering $>$ is simply a rule which allows us to unambiguously 
compare any two monomials in the variables and say which one is higher in a
list of all monomials. For example the Lexicographic ordering 
with respect to the variables $a,b,c$ just says that monomials are ordered,
firstly according to the power of $a$ they contain (highest first), then 
according to the power of $b$ and finally that of $c$. So, for
example, $a^2 b c$ would be ordered higher than $a b^2 c^4$.

The reduction process of polynomial $A$ relative to polynomial $C$ is
then simply 
as follows. 
We subtract some (possibly monomial) multiple of $C$ from $A$ in such a 
manner as to cancel $A$'s leading term with respect to the ordering
$>$. If the
leading term can not be canceled in this way $A$ is simply left alone.

The $S$ polynomial of two polynomials $A$ and $B$ is simply given as follows.
Multiply $A$ and $B$ by the lowest degree monomials possible so that
the leading 
terms of the two results, $A'$ and $B'$, become equal. One then simply
subtracts one from the other, so that the leading terms cancel: $S= A' - B'$.

Gr\"obner bases have many uses, some of which we shall encounter later in 
these appendices. One particularly useful feature of these sets of polynomials
is that the reduction of any polynomial with respect to ${\cal G}$ does not 
depend upon the order in which we use the polynomials therein in the
reduction procedure. Another vital property is that given a monomial
ordering, the Gr\"obner basis (reduced with respect to itself)
is unique for any given ideal.
Unfortunately, one of the biggest hurdles in computational algebraic
geometry is that the algorithm for determining the Gr\"obner basis
can be very intensive.
\section{Primary Decomposition Algorithms}
In this section, we discuss in a little more detail the key algorithm
used throughout the paper.
There are now several primary decomposition routines available
\cite{GTZ,EHV,SY},
many of which are implemented in algebra systems such as
\cite{mac,sing}. We make
extensive use of the algorithm due to Gianni, Trager, and Zacharias
(GTZ) \cite{GTZ}
in this paper and so we shall now give a brief description of the basics of 
the algorithm's workings, following closely such texts as
\cite{DGPcomp,DL}.

The GTZ algorithm is built around the same splitting principle as was used 
in \sref{saturationexp};
that is, if $(I:f^{\infty})=(I:f^l)$ for some $l$, then
\bea
I= (I:f^{\infty}) \cap \gen{I,f^l} \ .
\eea
Given this fact the GTZ algorithm works by specifying the polynomials $f$
and by reducing the primary decomposition
of an ideal of dimension $d$ to a problem involving primary decompositions of 
zero dimensional ideals. An existing algorithm can then be employed to 
primary decompose the zero dimensional ideals.
We thus split our description into these two halves. First, we describe 
how the GTZ algorithm reduces everything to zero dimensional primary 
decompositions and finds a suitable $f$. 
Second, we give a brief discussion of how one obtains a primary
decomposition of a zero dimensional ideal.

\subsection{GTZ reduction} 
The first step is to reduce the $d$ dimensional decomposition problem 
to a $0$ dimensional one.
We start with an ideal $I$ in the ring $\IC[X_1,\ldots,X_n]$. 
First, choose a
maximal subset $Y= \{ Y_1,...,Y_d \}$ 
of the variables of the ring, $X=\{X_1, \ldots, X_n \}$, 
such that these variables are independent mod $I$. That is, $I \cap
\IC[Y_1,...,Y_d]= \{0\}$. Geometrically, $Y$ are the variables along
$L(I)$ and $X \setminus Y$, transverse. Thus, $d$ is the dimension 
of $I$. Now take the polynomials defining $I$ to be polynomials in
$I_{\IC(Y)[X \setminus Y]} \subset \IC(Y)[X \setminus Y]$. That is,
pretend that the $Y$ variables are coefficients. The ideal
$I_{\IC(Y)[X \setminus Y]}$, with all $Y$ variables in $I$ considered
as coefficients, is then zero dimensional.

Now, for our original ring $\IC[X]$, choose a monomial ordering $<$,
with $Y_i < X_j$ for all $i$ whenever $X_j \in (X \setminus Y)$. 
Take a Gr\"obner basis ${\cal G}$ of $I$ with respect to $<$. This is then 
also a Gr\"obner basis of $I_{\IC(Y)[X \setminus Y]}$, via restriction of
$<$ to $X \setminus Y$.
We are now in a position to isolate the $f$ which GTZ employ. We take $f$ to 
be the least common multiple of the leading coefficients of the polynomials 
in $\cal{G}$, with these polynomials taken to lie in
$\IC(Y)[ X \setminus Y]$.
The crucial observation is then the following:
\beq
\label{GTZ1}
I_{\IC(Y)[X \setminus Y]} \cap \IC[X] = (I:f^{\infty}) \ .
\eeq
Thus, of the two halves of the saturation decomposition 
$I= (I:f^{\infty}) \cap \gen{I,f^l}$,
the first factor can be addressed by a zero-dimensional
primary decomposition (to which we turn in the next subsection),
leaving us with only $I' = \gen{I,f^l}$, to deal with. 
We can then repeat the above process on $I'$, and iterate
until when there is nothing new in the second factor, i.e., when
a factor we already have lies within the starting 
point for the next iteration. 

\subsection{Zero dimensional Primary Decomposition}
Finding a primary decomposition of a zero dimensional ideal is relatively 
straightforward using Gr\"obner bases. Any zero dimensional ideal $I$ 
can be put in a so-called ``general position'' with respect to the 
lexicographical ordering induced from $X_1>...>X_n$. This is defined 
by the following properties:
\begin{itemize}
\item The primes $P(i)$ in the primary decomposition of $I$ have a reduced
Gr\"obner basis with respect to the same ordering of the form
\be
\{P(i)\} =
\{ X_1 - h_1(X_n), \ \ldots, \ X_{n-1} -h_{n-1}(X_n), \ h_n(X_n) \} \ .
\ee
Here, we have $h_i \in \IC[X_n]$, i.e., they 
are simply polynomials in $X_n$.
\item The ideals $P(i)$ are coprime.
In other words, the polynomials $h_i$ have as their greatest common
divisors just an element of the coefficient field $\IC$, viz., a constant.
\end{itemize}
Write $\cal{G}$ for a corresponding minimal Gr\"obner basis and 
define $\{ h \} = {\cal G} \cap \IC [ X_n ] $. 
There is then a theorem \cite{DGPcomp} which 
states that if $h=h_1^{l_1} ... h_f^{l_f}$ is the factorisation of $h$ into 
a product of powers of pairwise non-associated irreducible factors, then the
primary decomposition is just given by:
\be
I = \bigcap_{j=1}^f \gen{I,h_j^{l_j}} \ .
\ee
An example of how this theorem can be used to implement an appropriate 
algorithm can be found in \cite{DGPcomp}, as can various details.

%

\section{Sturm Queries and Real Roots}
One of the topics of primary importance within this paper is 
the discussion of finding  real roots of zero dimensional ideals.
We shall thus briefly describe some of the mathematical ideas involved in 
this appendix, following closely the excellent treatments of
\cite{realAG,tobis}.

To commence, a finite set of polynomials within $\IC[X_1,...,X_k]$ is zero 
dimensional iff any Gr\"obner basis of the associated ideal contains a 
polynomial with leading monomial $X_i^{d_i}$ for each $i \in [1,k]$.
Once a zero dimensional system has been identified one of the central 
notions in the study of its real roots is that of a {\bf Sturm query}. 
Let $P \in \IR[x]$ be a real polynomial and $Z$, a set of points. 
The Sturm query is given by the following expression:
\bea
SQ(P,Z) = \sharp \{x \in Z | P(x)>0\} - \sharp \{x \in Z| P(x) < 0\} \ .
\eea

If we had this function, then, for a zero-dimensional ideal $I$
of real polynomials, the number of real roots is simply $SQ(1, r(I))$,
where $r(I)$ is the (discrete) set of real roots for $I$.
Moreover, we can also test sign
conditions, another real algebro-geometric device which we use extensively
in the paper. In such calculations we wish to know the sign 
taken by a given polynomial $P$ evaluated at the elements of $r(I)$.
We note that, by definition, the following system of equations holds:
\be
\left( \ba{ccc} 1&1&1\\ 0&1&-1 \\ 0&1&1 \ea \right) \left( \ba{c}
\sharp\{ x \in r(I) | P=0)\} \\ \sharp\{x \in r(I) | P>0)\} \\  
\sharp\{x \in r(I) | P<0)\}\ea \right) = \left( \ba{c} SQ(1,r(I))\\
SQ(P,r(I))\\ SQ(P^2,r(I))  \ea \right) \ .
\ee
Once the Sturm queries are known, we can immediate solve for the
quantities $\sharp\{x \in r(I) | P=0, P>0, \mbox{ or } P<0\}$, which
are what we are after. One can also, in the same way, 
ask about the signs of lists of 
polynomials. This just involves the study of a bigger matrix equation.

Thus we see that, once we know how to algorithmically compute Sturm
queries, we 
can find the number of real roots of an ideal as well as the 
signs various polynomials take on those roots.
How then is a Sturm query obtained algorithmically? The starting point here is 
to notice that if $I$ is zero dimensional then the quotient ring $R_Q=R[X_1,
\ldots,X_n]/I$ is a finite-dimensional $R$-vector space $A$. We can 
imagine taking a basis consisting of functions which are 1 on one root and 
zero on all the others, with one such function in the basis for each root.
One can then obtain any function on
the roots by combining multiples of these basis elements 
in the correct manner.
We can define various linear maps on this space. One such map, $L_f:A 
\mapsto A$ can just be defined to be multiplication within $R_Q$ by a function
$f$. One can also consider bilinear maps $H_g: A \times A \mapsto R$ defined by
$H_g(f_1,f_2)= \textnormal{Trace}(L_{f_1 f_2 g})$. 
Clearly the matrix associated to $H_g$ in some 
basis for $A$ is symmetric. 

A theorem due to Hermite states that the Sturm 
query $SQ(g,r(I))$ is simply given by the signature of this symmetric matrix.
This is, in fact, intuitively obvious when thinking in terms of the basis 
described above.
This matrix can be obtained algorithmically using 
Gr\"obner bases \cite{realAG}.
Algorithmically the signature of symmetric matrices is easy to
find. All of the 
eigenvalues of a symmetric matrix are real and are given by the roots of its 
characteristic polynomial. The number of positive roots is then
determined by essentially Descartes' law of signs (or its generalisation,
the Budan-Fourier theorem) \cite{realAG}, i.e., by examining
the signs of the coefficients of the characteristic polynomial.

The methods describe above are not necessarily the fastest way to obtain the 
results required, particularly the number of real roots \cite{realAG,tobis}.
They are however the 
simplest to understand. The reader interested in further details of these 
kinds of calculations is referred to \cite{realAG,tobis}. 
From a practical stand point,
all of the algorithms concerned with real roots which we require have been 
implemented in \cite{sing} by Tobis \cite{tobis}.


\end{document}